\documentclass[conference]{IEEEtran}
\IEEEoverridecommandlockouts

\usepackage{cite}
\usepackage{amsmath,amssymb,amsfonts}
\usepackage{algorithmic}
\usepackage{graphicx}
\usepackage{textcomp}
\usepackage{xcolor}
\def\BibTeX{{\rm B\kern-.05em{\sc i\kern-.025em b}\kern-.08em
    T\kern-.1667em\lower.7ex\hbox{E}\kern-.125emX}}

\usepackage{hyperref}
\usepackage{url}
\usepackage{icomma}
\usepackage{soul}
\usepackage{subcaption}
\usepackage{graphicx}
\usepackage{multirow}
\usepackage{multicol}
\usepackage{caption}
\usepackage{algorithm}
\usepackage{algorithmic}

\usepackage{amsmath}
\usepackage{amssymb}
\usepackage{mathtools}
\usepackage{amsthm}
\usepackage{wrapfig}
\usepackage{booktabs} 
\usepackage{enumitem}

\usepackage{pdfpages}

\usepackage{dblfloatfix}
\usepackage{indentfirst}

\theoremstyle{plain}
\newtheorem{theorem}{Theorem}[section]

\theoremstyle{definition}
\newtheorem{definition}[theorem]{Definition}

\theoremstyle{remark}

\begin{document}

\title{Privacy Vulnerabilities in \\Marginals-based Synthetic Data}


\author{\IEEEauthorblockN{Steven Golob\IEEEauthorrefmark{1},
Sikha Pentyala\IEEEauthorrefmark{1},
Anuar Maratkhan\IEEEauthorrefmark{1}, 
Martine De Cock\IEEEauthorrefmark{1}\IEEEauthorrefmark{2}
}
\IEEEauthorblockA{\IEEEauthorrefmark{1}School of Engineering and Technology,
University of Washington Tacoma, USA}
\IEEEauthorblockA{\IEEEauthorrefmark{2}Department of Applied Mathematics, Computer Science, and Statistics, Ghent University, Belgium}
\thanks{Corresponding author: Steven Golob (email: golobs@uw.edu).}}


\maketitle
\thispagestyle{plain}
\pagestyle{plain}

\begin{abstract}

When acting as a privacy-enhancing technology, synthetic data generation (SDG) aims to maintain a resemblance to the real data while excluding personally-identifiable information. Many SDG algorithms provide robust differential privacy (DP) guarantees to this end. However, we show that the strongest class of SDG algorithms--those that preserve \textit{marginal probabilities}, or similar statistics, from the underlying data--leak information about individuals that can be recovered more efficiently than previously understood. We demonstrate this by presenting a novel membership inference attack, MAMA-MIA, and evaluate it against three seminal DP SDG algorithms: MST, PrivBayes, and Private-GSD. MAMA-MIA leverages knowledge of which SDG algorithm was used, allowing it to learn information about the hidden data more accurately, and orders-of-magnitude faster, than other leading attacks. We use MAMA-MIA to lend insight into existing SDG vulnerabilities. Our approach went on to win the first SNAKE (SaNitization Algorithm under attacK ... $\varepsilon$) competition. 

\end{abstract}


%
%

\section{Introduction}

Access to quality data is essential for machine learning applications. Yet recent laws like the GDPR\footnote{European General Data Protection Regulation, adopted in 2016, 
\url{https://gdpr-info.eu/}} 
in Europe and HIPAA\footnote{
\url{https://www.cdc.gov/phlp/publications/topic/hipaa.html}} in the U.S.,\footnote{Additionally, President Biden's AI Bill of Rights and Executive Order on AI \cite{biden2023executive} lay the groundwork for greater regulations being enacted, \url{https://www.whitehouse.gov/ostp/ai-bill-of-rights/}} 
place stringent, and well-founded protections on the treatment and trading of personal data.
An increasingly prevalent way to circumvent mishandling personal data is to instead use synthetic data. The aim of synthetic data generation (SDG) is to create data that both resembles the real data in some meaningful way while also hiding any personally-identifiable information. 
Striking a fair balance between these two imperatives is nontrivial; a whole host of SDG techniques have been developed recently in academia \cite{hu2023sok}, and are being deployed in industry to do so.\footnote{\url{https://mostly.ai}, \url{https://synthesis.ai}, 
\url{https://gretel.ai}, etc.}

SDG techniques with built-in privacy assurances have however been shown to leak private information from the training data \cite{chen2020gan, hayes2017logan, jordon2021hide}. One fundamental way of showing leakages is to conduct a \textit{membership inference attack} (MIA) on the synthetic data, the goal being to glean which records were present in the hidden training data \cite{houssiau2022tapas,van2023membership}. Practically speaking, 
a successful MIA on cancer-related synthetic data or an AI model could allow an attacker to determine whether an individual's data was included in the training dataset. This could indirectly reveal that the individual had cancer or aid the attacker in performing more advanced attacks to infer aspects of their medical history.

Our contribution is a novel membership inference attack, \textit{MAMA-MIA} (MArginal Measurement Aggregation-based Membership Inference Attack) on synthetic data. MAMA-MIA is designed to expose weaknesses in data generated by ``marginals-based'' SDG, a class of SDG that has been shown to produce the \textit{highest quality} synthetic tabular data \cite{tao2021benchmarking,pereira2023assessment}. 
Therefore, the acclaim given to marginals-based SDG should be accompanied with due caution. 

The strength of our attack is that it is simple and efficient. While other state-of-the-art MIAs require extensive computation of ``shadow'' synthetic datasets as a way of understanding the generator's behavior, MAMA-MIA performs \textit{minimal} shadow modelling. MAMA-MIA's ability to detect information from the hidden data is at par with, and in some cases exceeds, that of leading MIAs while being computationally much more lightweight. We demonstrate this by using MAMA-MIA to attack data generated 
using superior marginals-based SDG algorithms: MST \cite{mckenna2021winning}; PrivBayes \cite{zhang2017privbayes}; and Private-GSD \cite{liu2023generating}. Importantly, each of these algorithms provides robust differential privacy guarantees. 

The key to our attack's performance is in leveraging knowledge of which SDG algorithm was used. The fact that MAMA-MIA achieves a similar attack success rate compared to leading MIAs at often only a fraction of the computational cost is significant and may even have legal implications in determining whether the generated data constitutes anonymous data. Indeed, burgeoning privacy laws take the practical feasibility of attacks into account, including ``the costs and the amount of time required for identification'' \cite{GDPR2016a}.


Lastly, our membership inference heuristic and experiments are motivated by our participation in the SNAKE 
(SaNitization Algorithm under attacK ...$\varepsilon$) 
Challenge 
\cite{allard2023snake}, where our approach won
first place. 

(This paper is an extension of our non-archival workshop paper, detailing earlier MAMA-MIA experiments on MST and PrivBayes.)

%
%

\vspace{10pt}
\section{Preliminaries and related work}\label{sec:related_works}

\subsection{Synthetic data generation}\label{sec:sdg}
Many types of SDG algorithms have been developed to generate data in different ways. One prominent class of SDG employs generative adversarial networks (GANs). 
Algorithms using this paradigm perform quite well when generating images (e.g.~\cite{torkzadehmahani2019dp,xie2018differentially}). However, they tend not to produce high quality \textit{tabular} data \cite{pereira2023assessment,tao2021benchmarking}. 
The other main approach to SDG is to instead measure \textit{statistical} properties from the training data $D_{train}$ and preserve them in the synthetic data $D_{synth}$. 
Specifically, we focus on privacy-preserving ``marginals-based'' SDG algorithms, i.e.~those that maintain marginal probability distributions to keep this likeness.\footnote{It is worth briefly mentioning that similar statistics-preserving SDG algorithms rely \textit{indirectly} on marginals, via copulas to capture the dependencies between features. But we don't study copulas algorithms here. Furthermore, some algorithms might be classified as projection-based \cite{hu2023sok}. However, many of these still operate by measuring marginals \cite{aydore2021differentially, vietri2022private} and using projection to preserve them.} 
When it comes to tabular data, marginals-based SDG techniques are dominant in achieving the highest utility for machine learning applications while maintaining privacy \cite{pereira2023assessment,tao2021benchmarking}. 
Naturally, a higher level of scrutiny must accompany their acclaim.
New research has shown that diffusion-based generation of tabular data can produce data with impressive fidelity to the training data \cite{kotelnikov2023tabddpm, zhang2023mixed,truda2023generating, duDiffusion}. However, as of this writing, most of them are not differentially private, and those that are have not been compared against differentially private marginals-based SDG.

In this paper, we use our novel MAMA-MIA attack to investigate vulnerabilities of three state-of-the-art SDG algorithms that employ marginals in some way: MST \cite{mckenna2021winning}, PrivBayes \cite{zhang2017privbayes}, Private-GSD \cite{liu2023generating} (and we include an investigation on RAP \cite{aydore2021differentially} in the appendix). What is notable about these algorithms, unlike several popular others, is that they each provide robust differential privacy (DP) \cite{dwork2014algorithmic} guarantees to protect the training data. So to maximize the inherently-conflicting aims of achieving high quality and strong privacy, these algorithms use marginals and apply DP, respectively. Descriptions of MST, PrivBayes, RAP, and Private-GSD are included in Section ~\ref{sec:MAMAMIAtailored}. 

\vspace{7pt}\noindent{\textbf{Marginal probability distributions}}\label{sec:marginals}
Treating the features in a tabular dataset $D$ as random variables, a $k$-way marginal is a marginal probability distribution over a subset of $k$ features.
Let $\mathbf{V} = (V_1,V_2,\ldots,V_k)$ be a list of $k$ features (variables), and $\mathbf{v} = (v_1,v_2,\ldots,v_k)$ a list of values that the variables in $\mathbf{V}$ can assume. Furthermore, for $d \in D$, let $\mathbf{V}(d)$ be the values in instance $d$ for the variables in $\mathbf{V}$. We estimate the $k$-way marginal $P(\mathbf{V})$ over $D$ as 
\begin{equation}
\hat{P}_D(\mathbf{v}) = \frac{|\{d \,| \,d \in D \wedge \mathbf{V}(d) = v\}| }{|D|}
\end{equation}
Similarly, for a set of ``parent'' features $\mathbf{V} = (V_1,V_2,\ldots,V_k)$ and a ``child'' feature $V_c$, we denote the conditional probability distribution as $P(V_c \mid \mathbf{V})$ and estimate it over $D$ as 
\begin{equation}
\hskip -8pt
\hat{P}_D(v_c \mid \mathbf{v}) =
\frac{|\{d\, |\, d \in D \wedge V_c(d) = v_c\ \wedge \mathbf{V}(d) = \mathbf{v}\}|}
{|\{d\, |\, d \in D \wedge \mathbf{V}(d) = \mathbf{v}\}| }
\end{equation}

For scalability reasons, SDG algorithms typically use a number of $k$-way marginals for a relatively small value of $k$ to approximate the joint probability distribution of $D_{train}$.

\vspace{7pt}\noindent{\textbf{Differential privacy (DP)}}
The well-known definition of DP \cite{dwork2014algorithmic} states that a randomized algorithm $\mathcal{A}$ with range $O$ is $\varepsilon$-DP if for any two adjacent datasets $D_1$ and $D_2$ (datasets that differ in one entry)
\begin{equation}
Pr[\mathcal{A}(D_1) \in O] \leq e^\varepsilon  Pr[\mathcal{A}(D_2) \in O]
\end{equation}
In the context of this paper, $\mathcal{A}$ is an SDG algorithm, and $\mathcal{A}(D_1)$ and $\mathcal{A}(D_2)$ are the synthetic datasets obtained when applying $\mathcal{A}$ to real datasets $D_1$ and $D_2$. 
DP ensures that the inclusion or exclusion of any entry in the real dataset is obscured, in the sense that any output (synthetic dataset) obtained from  computations over the real dataset would have been similarly likely to be reached whether the entry was present in the dataset or not. Differentially private SDG algorithms are, in other words, \textit{designed} to withstand membership inference attacks. The privacy guarantees of course depend on the privacy budget $\varepsilon \geq 0$, with smaller values indicating stronger privacy guarantees. 



\begin{figure}
\centering
\includegraphics[width=.45\textwidth]{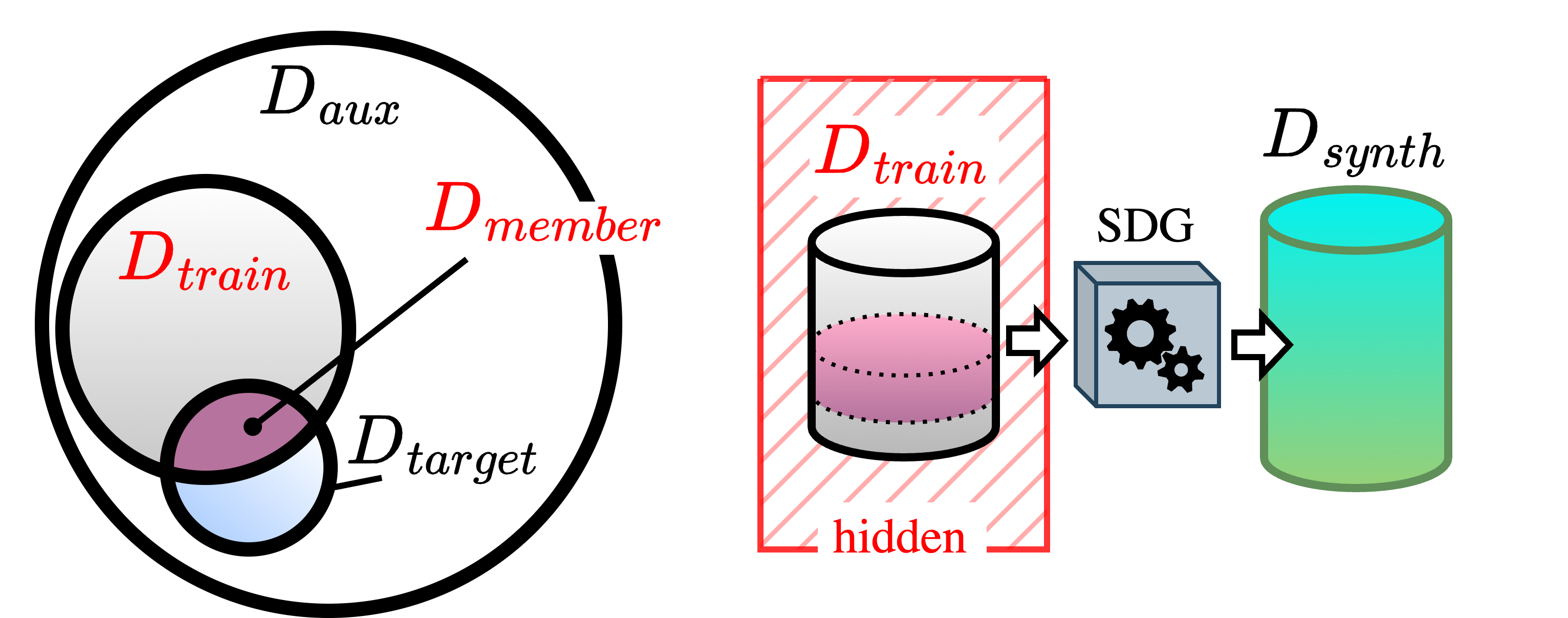}
\caption{Membership Inference Overview. The SDG algorithm takes $D_{train}$ as input, and produces $D_{synth}$. While an attacker has access to $D_{aux}$ and $D_{synth}$, $D_{train}$ remains hidden. The goal of the attack is to detect which records in $D_{target}$ are included in $D_{train}$.} 
\label{fig:MIA_diagram}
\end{figure}

\subsection{Membership inference}\label{sec:mia}

As illustrated in Figure \ref{fig:MIA_diagram}, a synthetic data generator is trained on real data $D_{train}$ and subsequently used to generate synthetic data $D_{synth}$. The goal of a membership inference attack (MIA) is to determine whether a given record was used to train the generator that produced the synthetic records without looking at the training data. We refer to the data records on which we make such inferences as ``targets'', $D_{target}$. MIAs can be conducted in other contexts too, such as on the output of a machine learning model instead of a synthetic dataset \cite{7958568}. But in this paper, we focus on the former, known as ``Inference-on-Synthetic'' attacks \cite{houssiau2022tapas}. 

We focus on two types of MIAs, namely ``single'' membership inference (MI) and ``set'' MI. Single MI refers to inferring the membership of individual records whereas in set MI, a target is a set of records (e. g. records belonging to individuals in the same household), and is either entirely present in $D_{train}$ or not.

\vspace{7pt}\noindent{\textbf{Threat model}}\label{sec:threat_model}
Each MIA makes assumptions about what an adversary has access to, known as the ``threat model''. Some MIAs use a threat model where an adversary knows little. Others grant a generous amount of knowledge to an adversary––the motivation being to drive the development of stronger privacy protections against stronger opponents. Our attack follows the latter inspiration. In our threat model, we use the ``auxiliary data assumption'', which admits knowledge of the population data to the adversary \cite{houssiau2022tapas}. This is a generally accepted assumption \cite{van2023membership, allard2023snake}, though is argued against in \cite{guepin2023synthetic}.

We also assume ``black-box knowledge of the generator'', which discloses the SDG algorithm used and its hyperparameters. A threat model using this assumption is in alignment with Kerckhoffs’s principle that the security of a system shouldn't rely on the secrecy of the algorithm, and is argued for in \cite{allard2023snake}, and in \cite{houssiau2022tapas} which states that this is the ``most common assumption, as it is consistent with good security practices.'' 

An even stronger threat model could grant ``white-box'' knowledge, where an adversary knows the internal randomness (seed) of the target generator, and its weights learned during training on the hidden data. But we do not use this liberal assumption in our attack. There also is a distinction between having black-box knowledge of the generator to create a synthetic dataset from one's own training data, not necessarily knowing the generator's implementation (a slightly weaker assumption, used in \cite{stadler2022synthetic}), and knowing its implementation (a stronger assumption, and used in \cite{allard2023snake}). We assume the latter here, even though most MIAs do not take advantage of this added information.

\subsection{Related MIAs}\label{sec:related_mias}

Like SDG techniques, there are different \textit{types} of MIAs that focus on different tasks and use different threat models. Here, we focus on ``Inference-on-Synthetic'' attacks \cite{houssiau2022tapas}.

References \cite{carlini2022membership} and \cite{stadler2022synthetic} describe attacks that rely on exhaustive ``shadow-modelling'' to learn about the hidden data. The idea behind these is to simulate the SDG algorithm on many random samples of the population data, thereby generating many ``shadow datasets'', $\hat{D}_{synth}$. By generating many shadow synthetic datasets, it may be possible to notice differences in those that were trained using the target (i.e. the example the attacker is interested in) versus those that were not, informing whether or not the target was a member of the \textit{actual} hidden training data $D_{train}$. 
Both these attacks use some kind of access to the model -- either as prior knowledge of the SDG model in \cite{stadler2022synthetic} or as query access to the model in \cite{carlini2022membership} -- to conduct the shadow modelling using one's own training data. And both works show superior inference accuracy using this method. However, it tends to require generating several hundred shadow datasets, using scads of compute resources.

An attack that uses a weaker threat model is proposed by \cite{guepin2023synthetic}. In this work, a methodology for removing the auxiliary data assumption is studied (when both including and excluding the black-box knowledge of the generator assumption), and is shown to yield strong accuracies, but is still categorically less accurate than when knowledge of the auxiliary data is assumed.

The TAPAS library\footnote{\url{https://tapas-privacy.readthedocs.io/en/latest/index.html}} 
provides a framework for several types of MIAs based on different threat models \cite{houssiau2022tapas}, including shadow modelling such as ``Groundhog'' attacks \cite{stadler2022synthetic}, distance-based attacks \cite{lu2019empirical, yale2019assessing},
probability estimation attacks \cite{jordon2022synthetic, hilprecht2019monte}, and others. In both their analysis and in \cite{meeus2023achilles}, shadow modelling attacks using random queries yield the highest accuracy when using the auxiliary data and black-box knowledge of the generator assumptions to generate shadow datasets. So in this paper, we provide a comparison of our attack with this leading MIA. (Meeus et al. find even greater success by smartly choosing \textit{which} target records are most susceptible to attack \cite{meeus2023achilles}. We do not compare our work against this specialized approach, since ours is less restricted, and \textit{randomly samples} which targets to attack.)

We also conduct a comparison of our attack against the DOMIAS MIA \cite{van2023membership}. DOMIAS is highly relevant because, in addition to making clever use of the auxiliary data assumption (expounded upon in Section \ref{sec:method}), it outperforms several other MIAs \cite{chen2020gan, hayes2017logan, hilprecht2019monte}. But it does \textit{not} use the black-box knowledge assumption, and is therefore generic to SDG algorithms.

A more detailed description of the MIAs we compare with in this work are given in Appendix \ref{sec:related_MIAs}.

\subsection{Problem description}\label{sec:problem}

We set up the problem as follows (and as shown in Figure \ref{fig:MIA_diagram}). Given a population, or ``auxiliary'', tabular dataset $D_{aux}$, given an $\varepsilon$-DP SDG algorithm $\mathcal{A}$ and a synthetic dataset $D_{synth}$ produced by running $\mathcal{A}$ on training dataset $D_{train} \subset D_{aux}$ (i.e. $\mathcal{A}(D_{train}) = D_{synth}$), and given some subset of ``targets'' from the population data $D_{target} \subset D_{aux}$, the goal is to determine which of $D_{target}$ were ``members'' of the training data, $D_{member} = D_{target} \cap D_{train}$. Always, $D_{train}$ is hidden.

\vspace{10pt}
\section{MAMA-MIA: the MArginals Measurment Aggregation-based Membership Inference Attack}\label{sec:method}

\begin{figure}
\centering
\includegraphics[width=8.4cm]{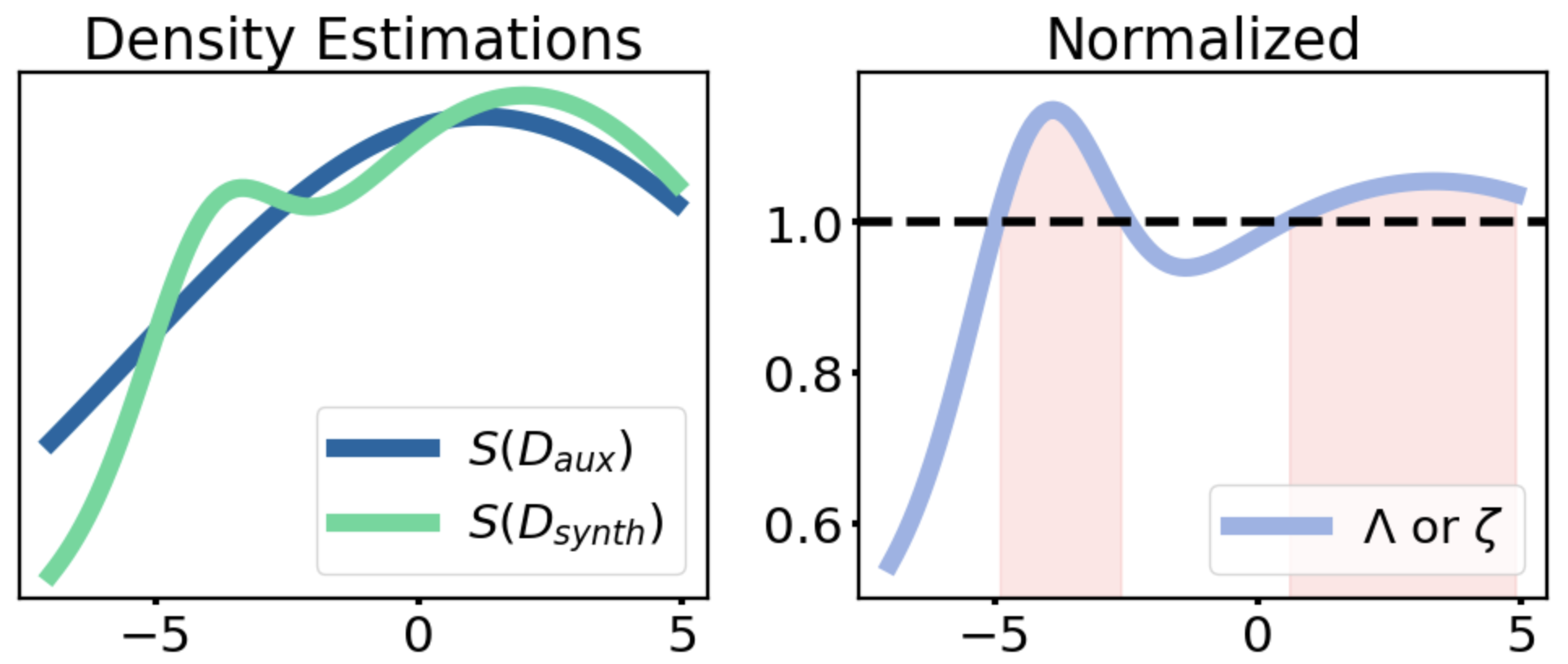}
\caption{A simple visualization of how DOMIAS and MAMA-MIA detect overfitting. $S$ gives some density estimation of $D_{synth}$ and $D_{aux}$ (left). Normalizing $S(D_{synth})$ by $S(D_{aux})$ exposes overfitting on $D_{train}$ (right).} 
\label{fig:domias_example}
\end{figure}

Our attack, MAMA-MIA, uses the auxiliary data assumption similar to DOMIAS, with the addition of leveraging the black-box knowledge of the generator used. This allows us to substantially increase inference accuracy. Specifically, both our attack and DOMIAS measure some density estimation $S$ on the synthetic data, and do the same for the auxiliary data. Then they normalize $S(D_{synth})$ by $S(D_{aux})$, which will expose where there are higher and lower densities in the synthetic data relative to the auxiliary data, which intuitively should expose the density characteristics of the training data if the generator was at all overfit to it. Without the auxiliary data assumption, it is more difficult to contextualize observations on $D_{synth}$ and detect overfitting \cite{guepin2023synthetic}. 

A simple illustration of this idea is given in Figure \ref{fig:domias_example} where, in the second graph, areas in red suggest where there might be a high concentration of training data values. (This idea is discussed in depth in \cite{van2023membership}.) DOMIAS uses \textit{generic} density estimators to this end, such as KDE \cite{scott2015multivariate} and BNAF \cite{de2020block}. Then it simply computes $\Lambda = S(D_{synth}) / S(D_{aux})$ for each target.\footnote{The threshold for inferring membership is when $\Lambda > 1.0$. But it needn't be. Normalization allows for any threshold to identify density concentrations meaningfully.} 

MAMA-MIA constructs a novel density estimation $\zeta$ in place of $\Lambda$. $\zeta$ incorporates the normalization idea from DOMIAS, but is tailored to which marginals-based algorithm was used, and therefore has a much stronger ability to detect characteristics of the training data. Applicable to any marginals-based SDG algorithm, the heuristic behind MAMA-MIA is to:

\begin{description}[leftmargin=10pt,itemsep=2pt,topsep=3pt]
    \item[Step 1] \textit{identify} how the SDG algorithm chooses which marginal measurements to take on $D_{train}$ (we will refer to the chosen measurements as ``focal-points'')
    \item[Step 2] \textit{simulate} the SDG algorithm several times (known as ``shadow modelling''), recording the frequency at which focal-points are chosen
    \item[Step 3] \textit{aggregate} the frequently-chosen focal-points into a density estimation $\zeta$, a replacement of $\Lambda$ in DOMIAS.
\end{description}


\begin{definition} 
We refer to a \textit{focal-point} (FP) as any statistical measurement taken on a dataset $D$ by the SDG algorithm. When the measurement is a marginal, the focal-point $\mathbf{F}$ is defined by the list of features 
$\mathbf{V}=(V_1,V_2,\ldots,V_k)$, i.e.~$\mathbf{F}=\mathbf{V}$. When the measurement is a conditional, it is defined by the child feature $V_c$ and the list of parent features $\mathbf{V}=(V_1,V_2,\ldots,V_k)$, i.e.~$\mathbf{F} = V_c \mid \mathbf{V}$. For simplicity, in both cases we use $\hat{P}_D(\mathbf{F})$ to denote the estimate of the marginal or conditional probability distribution respectively.
\end{definition}

Each SDG algorithm discussed in this paper chooses differently which set of FPs $F$ it will use to measure $D_{train}$ and to preserve in $D_{synth}$, but each essentially makes its choices to maximize mutual information (or some variant thereof) between the marginals' features. These focal-points are highly relevant to an adversary because they show what the generators measured in $D_{train}$! Because of each algorithm's application of DP noise, this FP selection process is nondeterministic. This is where we make use of shadow modelling.

\begin{algorithm}[t]
\caption{MAMA-MIA}
\label{alg:customMIA}
\textbf{Input}: $D_{synth}$ the synthetic dataset, $D_{aux}$ the auxiliary dataset, $\mathcal{F}$ a set of focal-points, $w_\mathbf{F}$ the weight of $\mathbf{F} \in \mathcal{F}$, $D_{target}$ the target records. \\
\textbf{Output}: $\zeta$ the list of density estimations of ($D_{synth} / D_{aux}$) for each $t \in D_{target}$
\begin{algorithmic}[1] 
\STATE Let $\mathcal{P}_{D_{synth}} = 
\{\hat{P}_{D_{synth}}(\mathbf{F}) \mid \mathbf{F} \in \mathcal{F}\}$ be the measurements of focal-points in $\mathcal{F}$ from $D_{synth}$  
\STATE Let $\mathcal{P}_{D_{aux}} = 
\{\hat{P}_{D_{aux}}(\mathbf{F}) \mid \mathbf{F} \in \mathcal{F}\}$ be the measurements of focal-points in $\mathcal{F}$ from $D_{aux}$ 
\STATE Let $\zeta$ = [\,]
\FOR{$t \in D_{target}$}
\STATE Let $\zeta[t] = \displaystyle\sum_{\mathbf{F}^{} \in \mathcal{F}} w_{\mathbf{F}} \cdot \hat{P}_{D_{synth}}(\mathbf{F}(t)) / \hat{P}_{D_{aux}}(\mathbf{F}(t))$
\ENDFOR
\STATE \textbf{return} $\zeta$
\end{algorithmic}
\end{algorithm}

\subsection{Shadow modelling in MAMA-MIA}\label{subsec:shadow}

In step 2, we simulate the creation of $D_{synth}$ by running the same SDG algorithm several times, using the same privacy budget $\varepsilon$ and training conditions. Since we do not know the true ${D}_{train}$, we use random samples of $D_{aux}$ as $\hat{D}_{train}$. From this process, we can record how frequently each FP is chosen, thereby developing a confidence of which FPs were likely chosen when generating the actual $D_{synth}$. For example, if we notice that MST chooses to measure the 2-way marginal probability of feature-pair $\{$`Age', `Income'$\}$ 48 times out of 50 shadows runs, we can say that it measured this marginal from ${D}_{train}$ and maintained it in $D_{synth}$ with high likelihood. Our shadow modelling step involves obtaining an implementation of the SDG algorithm, and modifying it to record the focal-points (step 1). This step requires substantially less shadow runs to develop a confidence than Groundhog attacks \cite{stadler2022synthetic,houssiau2022tapas}
(we found success with 50 runs). Furthermore, no exhaustive computation or storage is necessary on any $\hat{D}_{synth}$, and we even terminate the SDG process early once the FPs are determined! As we show in Section \ref{sec:results}, this reduces the runtime from days to minutes without sacrificing attack success rates.

\subsection{Estimating the densities $\zeta$ on marginals-based synthetic data}

Once the oft-chosen FPs are known, we can use them to build a density estimation of a dataset that captures information resembling how the SDG algorithm would capture or approximate that same dataset. So presumably, our density estimation of $D_{synth}$ would look like that of $D_{train}$. Furthermore, we can better distinguish the shape of $D_{train}$ by normalizing these FP measurements on $D_{synth}$ by the same measurements taken on the auxiliary data $D_{aux}$.

This custom density estimation $\zeta$ is depicted in Algorithm \ref{alg:customMIA}. Lines 1--2 are made unique to each SDG algorithm (see Section \ref{sec:MAMAMIAtailored}). The overall aggregation of focal-point measurements into $\zeta$ is the same for all. 

How this aggregation is done is discretionary. Our approach achieves success by summing focal-point measurements from $D_{synth}$ for a particular target's value, divided by the same measurement on $D_{aux}$ (line 5). We offer a very similar, alternative aggregation method in Appendix \ref{sec:alternative_zeta}. Since the true focal-points are chosen stochastically, we also weight each focal-point measurement by its frequency chosen during shadow modelling. This weighting is less important when $\varepsilon$ is large and the generator's focal-point choices become more deterministic. 

Once $\zeta$ is measured for the targets, we convert it to a probability of membership, $P \in [0, 1]$ by designing an activation function. Like DOMIAS and other works on MIAs, we omit detailing this step here, but provide a description in Appendix \ref{sec:activation}.

\vspace{7pt}\noindent{\textbf{Adaptation to set MI}}
For set MI, we first make inferences for each individual record in a target set, then simply take the average of those predictions. This aggregation method scored better than other approaches, such as trusting the most confident individual inference of the set, or other trivial weightings. We make note of this here because set MI was the objective in the SNAKE Challenge \cite{allard2023snake}.

\subsection{Adaptation of MAMA-MIA to each SDG algorithm}\label{sec:MAMAMIAtailored}

\noindent{\textbf{MAMA-MIA on MST}}
MST \cite{mckenna2021winning} builds a graphical approximation of the joint probability distribution of $D_{train}$ where the nodes are the features of $D_{train}$, the edges are the two-way marginal probabilities between two features, and the graph is an undirected tree. During synthesis, MST creates data samples in proportion with the probabilities of all of the marginals measured in the graph. 

The edges chosen by MST (i.e.~marginals) are the focal-points in our density estimation, so how the graph is constructed is most interesting to us. MST attempts to draw edges that create a maximum spanning tree (hence ``MST'') based on the mutual information of each feature-pair. But the attempt is inexact because of the differential privacy mechanisms applied to this decision process. Half of the privacy budget is spent on selecting the optimal edges, and half is spent on performing the marginal measurements themselves.

During one shadow modelling run, we will observe that the amount of FPs chosen is $l - 1$, where $l$ is the number of features in $D_{train}$, since that's how many edges are in a tree of $l$ nodes. In Algorithm \ref{alg:customMIA}, we input the counts of all focal-points $\mathcal{F}$ observed during shadow modelling as weights, $w_\mathbf{F}$ for $\mathbf{F} \in \mathcal{F}$. So the total of all of the weights $\sum_{\mathbf{F}^{} \in \mathcal{F}} w_{\mathbf{F}} = 50 \cdot (l - 1)$, with some FPs being observed more frequently than others.

Off the cuff, there is one defense MST can take; MST also provides the option to manually select $k$-way marginals for its estimation of the training data's probability distribution. These hand-picked marginals obviously cannot be determined through shadow-modelling, which only simulates default behavior, and so would weaken an adversary's ability to build a similar estimation.

\vspace{7pt}\noindent{\textbf{MAMA-MIA on PrivBayes}}
Our tailored MIA on PrivBayes synthetic data follows a similar approach. PrivBayes \cite{zhang2017privbayes} also estimates the probability distribution of $D_{train}$ by constructing a graph. Except that, while MST constructs an undirected tree, with exactly $n - 1$ edges by default, PrivBayes constructs a \textit{directed} graph, where edges represent important \textit{conditional} probabilities between ``child'' features and sets of ``parent'' features.

Like MST, it uses the conditionals-based graphical approximation in order to generate $D_{synth}$, making this approximation highly relevant to an adversary. Edges that yield high mutual information are preferred by PrivBayes, but how many parents are allowed in each conditional depends on $\varepsilon$.

Specifically, when $\varepsilon$ is small, PrivBayes reduces $k$, the maximum number of parents allowed in its graphical approximation. Intuitively, setting $k$ to be large can allow conditionals to more closely approximate the true probability distribution. However, it also means that the high-specificity conditionals are more susceptible to the DP-noise. Throttling back $k$ mitigates this. So entirely different conditionals are tended towards for different $\varepsilon$. We use these conditionals as our focal-points for PrivBayes, and predict which are chosen during shadow modelling, using the correct $\varepsilon$.

\vspace{7pt}\noindent{\textbf{MAMA-MIA on Private-GSD}}
Private-GSD is a novel approach to generating data, employing an evolutionary algorithm \cite{liu2023generating}. Like the other SDG algorithms, this one also chooses some set of marginals (2-way by default) over the one-hot encoded data, and then takes DP noisy measurements of these on the real data. Then, it preserves these measurements in a very different way. It instantiates a ``population'' of synthetic datasets (filled initially with random values). Then, during each ``generation'' (i.e. iteration), some elite subset of synthetic datasets from the population (those that yield the lowest error to the chosen marginal measurements) is carried over to the next generation. Also, the dataset from that generation with the least error is reproduced for the next generation with slight mutations and crossovers with other elite datasets. Over the course of thousands of generations, the population's datasets will yield lower and lower error with respect to the selected marginals. After some training condition is met, Private-GSD outputs the synthetic dataset with the least error. This algorithm is more memory and computationally exhaustive than MST \cite{mckenna2021winning} and PrivBayes \cite{zhang2017privbayes}, but comparable with RAP \cite{aydore2021differentially}.

MAMA-MIA finds success against Private-GSD following the same process as before. We first simulate Private-GSD on subsets $\hat{D}_{train}$ of $D_{aux}$ and record which FPs (i.e. marginals) are chosen each time. Across 50 runs, some FPs are chosen consistently and others are not. Once we have the FPs $\mathbf{F} \in \mathcal{F}$ and their respective frequencies $w_{\mathbf{F}}$, these can be inputted into $\zeta$ in Algorithm \ref{alg:customMIA}.

\vspace{7pt}\noindent{\textbf{A note on generalizability and efficiency}} 
Our density estimation approach can be tailored to similar SDG algorithms that estimate the probability distribution of a training dataset via marginals or other similar statistics, like conditionals or copulas. In Appendix section \ref{sec:MAMAMIAtailored}, for example, we explain how MAMA-MIA can be adapted to RAP, another recent marginals-based SDG algorithm, and provide experimental results for it.
Additionally, MAMA-MIA is highly efficient, and operates in linear time with respect to the size of $D_{synth}$ and $D_{aux}$ (we offer an asymptotic analysis in Appendix \ref{sec:efficiency}). Not only does it not require generating hundreds of shadow datasets, but MAMA-MIA also has the benefit of operating on many targets at once without the runtime being noticeably impacted. That is, it is not necessary to train a new classifier per target.

\begin{figure*}[ht!]
\centering
\includegraphics[width=\textwidth]{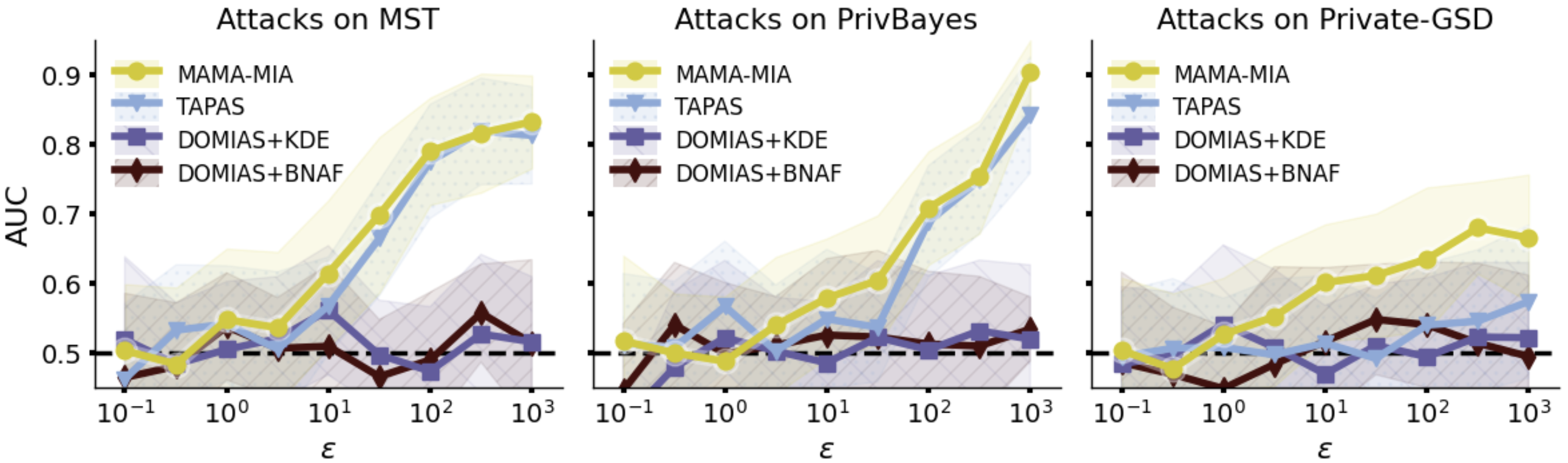}
\caption{\textbf{Experiment A} The accuracy of membership inference performed with MAMA-MIA is at par with, or higher than, the accuracy of state-of-the-art membership inference attacks TAPAS \cite{houssiau2022tapas} and DOMIAS with BNAF/KDE \cite{van2023membership}. The results are consistent regardless of the SDG algorithm used to generate the synthetic data, i.e.~MST \cite{mckenna2021winning}, PrivBayes \cite{zhang2017privbayes}, and Private-GSD \cite{liu2023generating}.} 
\label{fig:expA}
\end{figure*}

%
%

\vspace{10pt}
\section{Experiments and results}\label{sec:results}

We perform a series of experiments on MAMA-MIA to show its leading accuracy, and to demonstrate its \textit{far superior} efficiency compared to state-of-the-art MIAs -- TAPAS \cite{houssiau2022tapas} and DOMIAS \cite{van2023membership}. These experiments include attacking each SDG algorithm -- MST, PrivBayes, and Private-GSD -- under various configurations.

Each experiment is averaged over 30 trials.\footnote{We only perform 10 trials for DOMIAS-using-BNAF experiments due to its exorbitant computational cost. See Table \ref{tab:expA_runtimes}.} We record all runtimes, and we evaluate inference accuracy using AUC over our predictions for all of the targets. For each trial, a new training dataset $D_{train}$ is randomly sampled from the population data. To ensure consistency, this training dataset is shared among all SDG algorithms in the trial to produce a synthetic dataset $D_{synth}$. Subsequently, all MIAs attack the same $D_{synth}$ in the trial, for each SDG algorithm.

In \textbf{experiment A}, we compare the accuracies of our MAMA-MIA attack with other attacks across nine privacy levels $\varepsilon$. Then in \textbf{experiment B}, we demonstrate how the attacks perform across various \textit{size} configurations, ranging from 100 to 30,000+ records,
showing that MAMA-MIA handles larger datasets better. \textbf{Experiment C} is about sparing no expense in computational resources for state-of-the-art MIAs to show that, still, MAMA-MIA outperforms them while using only modest resources. \textbf{Additional experiments} include our strong results in the SNAKE Challenge \cite{allard2023snake}, where MAMA-MIA won first place. This, and  experiments A, B, and C are conducted on the SNAKE data (described below). Then we extend the above experiments to the very dissimilar California Housing dataset, and to when the training data does not overlap with the auxiliary data. We also provide a simple 
\textit{quality} measurement of the synthetic data produced to accompany the attack accuracies. And for those interested, we offer \textbf{extended results} in the appendix for further evidence of the robustness of our findings, including when MAMA-MIA attacks another recent marginals-based SDG algorithm, RAP \cite{aydore2021differentially}.

We ran all experiments with the default settings of the SDG algorithms and the MIAs, unless otherwise specified. (More hyperparameters are given in Appendix \ref{hyperparameters}.) All experiments are conducted on an AMD Ryzen Threadripper 3990X 64-Core Processor with 264GB of memory.


\vspace{7pt}\noindent{\textbf{A note on our TAPAS setup}}
The TAPAS attack is set up to generate shadow datasets for one target at a time. However, in our experiments, we attack a set of targets, as is done in \cite{allard2023snake}. So we modify the TAPAS library to make shadow modelling much more efficient. We do this by making sure that all targets are in a different 50\% of the shadow datasets, which allows us to reuse the shadow datasets in the attack for each target, thereby only requiring us to shadow model once, similar to the setup of \cite{carlini2022membership}.

While in \cite{houssiau2022tapas} random queries are fed into the random forest classifier at the heart of the TAPAS attack, we feed it all possible 3-way queries.\footnote{An ``$n$-way query'' is simply the count of rows in a dataset that match the target record's values over a set of $n$ columns.} In our investigation, using all queries either improved or did not affect the accuracy of TAPAS. This makes sense because a random forest will choose the best queries to use. Additionally, we explored using 1-way, 2-way, and 5-way queries. 3-way queries always yielded the highest accuracy.

Unless otherwise stated, we conduct all TAPAS attacks with 500 shadow datasets each.

\vspace{7pt}\noindent{\textbf{SNAKE dataset}}
We use the same dataset in our experiments that was used in the 2023 SNAKE competition \cite{allard2023snake}, referred to as the ``SNAKE'' dataset. It is comprised of three numerical features, two finite ordered features, and ten purely categorical features. It holds demographic information. To give a sense, the features include age, state of residency, number of children, marital status, ethnicity, gender, field of profession, weekly hours worked, income bracket, etc. Each record represents an individual, and individuals are grouped by a household identifier, which we use as record sets during set MI. This dataset has 201,279 records, containing 77,111 households (i.e. record sets). Of the households, there are 812 with at least 5 records.

\vspace{7pt}\noindent{\textbf{California Housing dataset}}
We also use \textit{sklearn}'s sample California Housing Dataset\footnote{\url{https://scikit-learn.org/stable/modules/generated/sklearn.datasets.fetch_california_housing.html}}, which has been used frequently in machine learning research. The data holds information on residential homes and households by district, collected as part of the 1990 U.S. Census. It contains nine continuously-valued features, and has a total of 20,640 records. It is useful for our purposes to compare against the more categorical-heavy SNAKE data. We create arbitrary groupings of 5 records to use as sets during set MI experiments. We also transform values from each column into twenty \textit{equal-depth} bins, to input into the SDG algorithms.

\vspace{7pt}\noindent{\textbf{Code base}} 
The MAMA-MIA scripts used to run these experiments can be found at 
\url{https://github.com/steveng9/SyntheticData_MIA}.

\subsection{Experiment A results}\label{sec:expA}

We compare the attack accuracies of MAMA-MIA with the leading attack from the TAPAS library (cfr.~Section \ref{sec:related_mias}), and DOMIAS (cfr.~Section \ref{sec:related_mias}) on all three SDG algorithms across nine privacy levels\footnote{Though not used in practice, we perform experiments on $\varepsilon$ up to 1000 to gain insight into the success rate of the MIAs on synthetic data that was generated with (almost) no differential privacy guarantees} $\varepsilon \in \{10^{i/2} | -2 \leq i \leq 6\}$, fixing $|D_{train}| = 1,000$ records of SNAKE data (see configuration (iii) 
in Table \ref{tab:dataset_sizes}).

The TAPAS attack uses 500 shadow datasets. We run it using all possible 3-way queries as features for its model, as well as all possible 5-way queries. Over the 15 features in the SNAKE data, this ends up being 455 queries, and 3,003 queries respectively. We present results for 3-way queries since it yielded better results.

Results are shown in Figure \ref{fig:expA}. The shaded regions depict +/- one standard deviation from the mean AUC. Our MAMA-MIA attack's accuracy is at par with or better than the other MIAs on data generated by all three SDG algorithms. Not surprisingly, the accuracy improves when the privacy budget $\epsilon$ is larger, i.e.~when the privacy guarantees offered by the DP SDG algorithms are lower. DOMIAS did not perform significantly above random guessing, neither with BNAF nor with KDE, the two density estimators used by Van Breugel at al.~\cite{van2023membership}.\footnote{There were settings in which DOMIAS performed better than random guessing, but still far worse than MAMA-MIA, and only when the datasets used were small (see Appendix Section \ref{sec:RAP}).} We also provide a tabled version of our full results, with standard deviations, as well as TPR / FPR curves when $\varepsilon = 10$ in the Appendix \ref{sec:extended_results}. 

But the striking advantage our MAMA-MIA technique has over its fellow MIAs is its efficiency. Shown in Table \ref{tab:expA_runtimes} are the average runtimes of each of the MIAs in experiment A. These times include the full attack pipeline on the set of targets; that is, both the shadow modelling stage and the inference stage of the MAMA-MIA attack, and the TAPAS attack. Recall from Section \ref{subsec:shadow}, we only shadow model 50 times for MAMA-MIA, where we terminate the SDG process early (as soon as the focal-points are determined). This is particularly relevant for SDG algorithms that are computationally intensive, like Private-GSD, where the TAPAS shadow modelling step entails 500 complete executions.\footnote{The times given are processor time, and so can be shortened with parallel processing.} For DOMIAS, there is no shadow modelling. But even so, when using BNAF, DOMIAS takes the longest (except for on Private-GSD).

\begin{figure*}
\centering
\includegraphics[width=\textwidth]{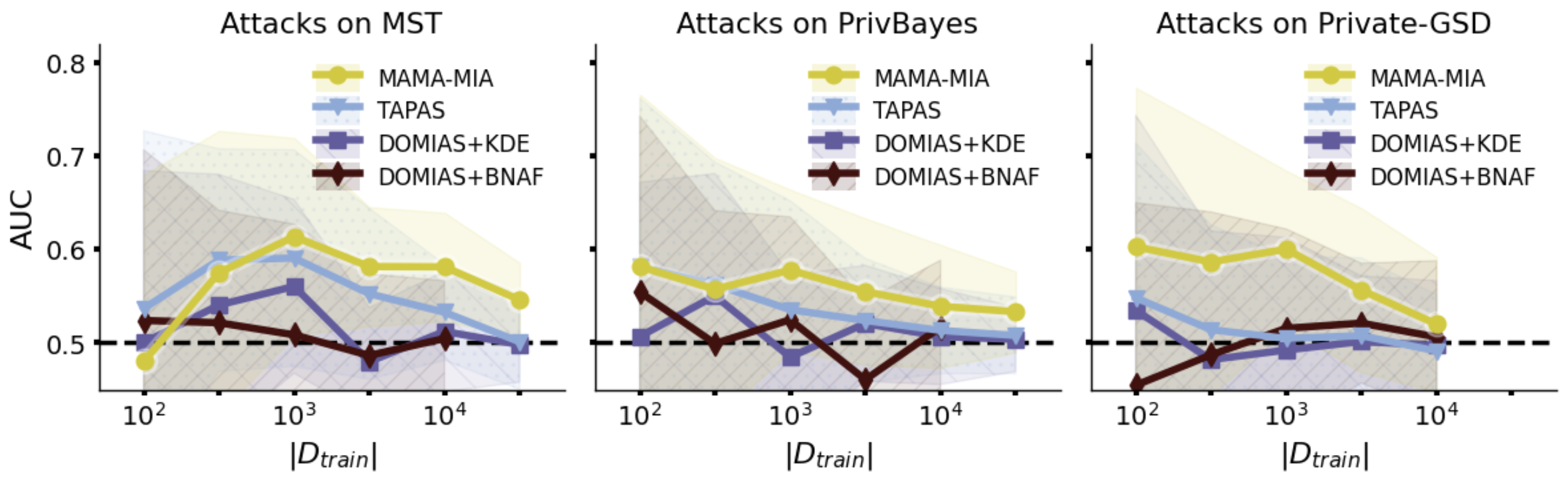}
\caption{\textbf{Experiment B} Trialing size configurations (i.) - (vi.) in Table \ref{tab:dataset_sizes} on the SNAKE data, when $\varepsilon = 10$. (We omit $|D_{train}| = 31,623$ results for Private-GSD and for the DOMIAS+BNAF attack due to computational limitations.)} 
\label{fig:expB}
\end{figure*}

\begin{figure*}[ht!]
\centering
\includegraphics[width=\textwidth]{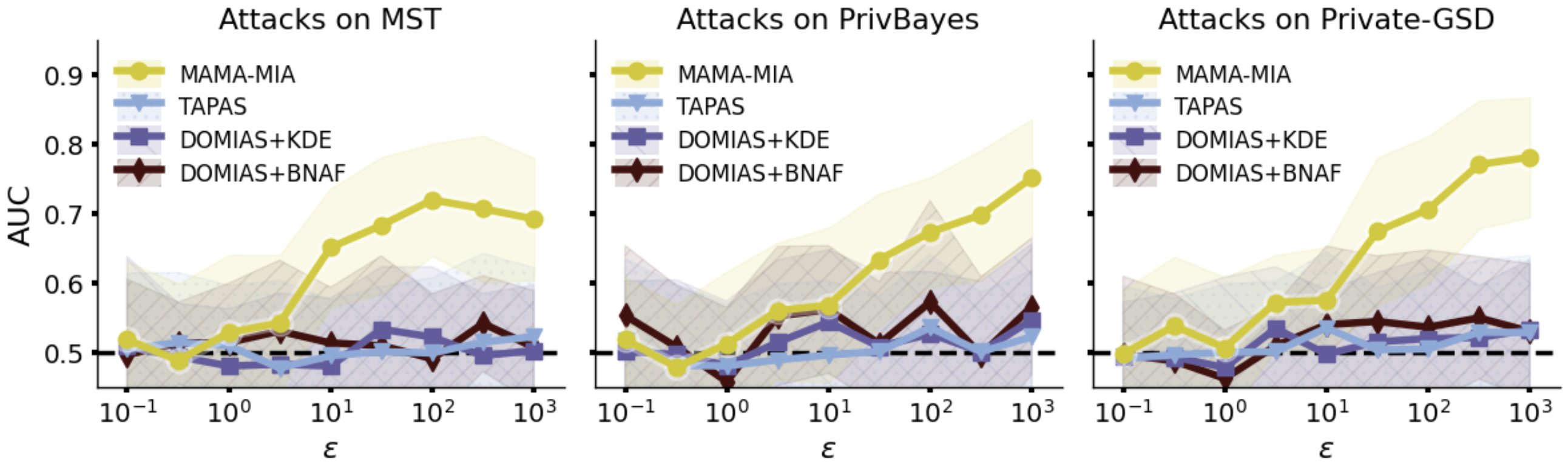}
\caption{Conducting membership inference attacks on the housing data.} 
\label{fig:expD_cali}
\end{figure*}

\vspace{7pt}
\begin{table}[!h]
\caption{Total pipeline runtimes for attacking all targets in one trial in \textbf{experiment A}, where $|D_{train}| = 1,000$ records of SNAKE data.}
\label{tab:expA_runtimes}
\centering
\begin{tabular}{c | c c c c}
\multicolumn{1}{c}{} & \multicolumn{4}{c}{\textbf{Attacks on MST}} \\
    \multirow{2}{*}{$\varepsilon$} & DOMIAS & DOMIAS & \multirow{2}{*}{TAPAS} & \multirow{2}{*}{MAMA-MIA} \\
    & +KDE & +BNAF &  &  \\
    \hline
    0.1 & 2.55 \textbf{sec} & 8.88 \textbf{hours} & 3.52 \textbf{hours} & 12.2 \textbf{min} \\
    1 & 2.53 \textbf{sec} & 8.92 \textbf{hours} & 4.02 \textbf{hours} & 13.2 \textbf{min} \\
    10 & 2.34 \textbf{sec} & 8.09 \textbf{hours} & 4.45 \textbf{hours} & 14.0 \textbf{min} \\
    100 & 1.88 \textbf{sec} & 8.96 \textbf{hours} & 4.81 \textbf{hours} & 14.6 \textbf{min} \\
    1000 & 2.76 \textbf{sec} & 8.82 \textbf{hours} & 4.73 \textbf{hours} & 14.7 \textbf{min} \\
    \\
    & & & & \\
    \multicolumn{1}{c}{} & \multicolumn{4}{c}{\textbf{Attacks on PrivBayes}} \\
    \multirow{2}{*}{$\varepsilon$} & DOMIAS & DOMIAS & \multirow{2}{*}{TAPAS} & \multirow{2}{*}{MAMA-MIA} \\
    & +KDE & +BNAF &  &  \\
    \hline
    0.1 & 1.61 \textbf{sec} & 8.74 \textbf{hours} & 3.42 \textbf{min} & 8.65 \textbf{sec} \\
    1 & 0.707 \textbf{sec} & 8.82 \textbf{hours} & 2.03 \textbf{min} & 14.2 \textbf{sec} \\
    10 & 2.80 \textbf{sec} & 8.22 \textbf{hours} & 3.28 \textbf{min} & 11.2 \textbf{sec} \\
    100 & 2.52 \textbf{sec} & 8.86 \textbf{hours} & 8.70 \textbf{min} & 29.1 \textbf{sec} \\
    1000 & 2.41 \textbf{sec} & 9.06 \textbf{hours} & 1.71 \textbf{hours} & 3.87 \textbf{min} \\
    \\
    & & & & \\
    \multicolumn{1}{c}{} & \multicolumn{4}{c}{\textbf{Attacks on Private-GSD}} \\
    \multirow{2}{*}{$\varepsilon$} & DOMIAS & DOMIAS & \multirow{2}{*}{TAPAS} & \multirow{2}{*}{MAMA-MIA} \\
    & +KDE & +BNAF &  &  \\
    \hline
    0.1 & 1.79 \textbf{sec} & 9.29 \textbf{hours} & 4.76 \textbf{days} & 1.99 \textbf{min}  \\
    1 & 1.66 \textbf{sec} & 9.02 \textbf{hours} & 9.10 \textbf{days} & 2.61 \textbf{min}  \\
    10 & 1.60 \textbf{sec} & 8.40 \textbf{hours} & 19.0 \textbf{days} & 6.41 \textbf{min}  \\
    100 & 0.903 \textbf{sec} & 8.74 \textbf{hours} & 35.6 \textbf{days} & 13.8 \textbf{min}  \\
    1000 & 1.46 \textbf{sec} & 9.23 \textbf{hours} & 41.4 \textbf{days} & 19.1 \textbf{min}  \\
\end{tabular}
\end{table}

\subsection{Experiment B results}\label{sec:expB}

This experiment, shown in Figure \ref{fig:expB}, is about demonstrating how much more \textit{consistently} strong our MIA is as the datasets become large, as opposed to the TAPAS MIA, which only appears to perform well on smaller dataset sizes. On the other hand, MAMA-MIA appears to far outperform all others at smaller sizes on Private-GSD. 

We fix $\varepsilon = 10$, as is done in \cite{houssiau2022tapas} and is generally considered reasonable \cite{nistdifferentialprivacy}. Again, we are using the SNAKE data, and TAPAS uses 500 shadow datasets. We report results for the size configurations\footnote{We omit the configuration where $|D_{train}| = 31,623$ for Private-GSD because of its hefty resource costs. Even only using 20,000 generations, Private-GSD generating only one of these datasets did not terminate after two days.} (i) - (vi) in Table \ref{tab:dataset_sizes}, where the sizes in each column are spaced evenly on a logarithmic scale. Notice that, in these configurations, not only do the training and synthetic dataset sizes vary; also we steadily decrease the \textit{proportion} of targets across configurations, starting at 10\% of the size of the training data, and going to 0.56\% of the training data size. 
As is done in \cite{allard2023snake}, we make the number of actual members $|D_{member}|$ half of the number of targets.

\begin{table}
\caption{Dataset size configurations used in \textbf{experiment B} and others. Each trial uses configurations entirely from one row.}
\centering
\small{
\begin{tabular}[b]{c| c c c c}
    & $|D_{train}|$ & $|D_{synth}|$ & $|D_{target}|$ & $|D_{member}|$ \\ 
   \hline
   i. & 100  &  100  &  10  &  5  \\
   ii. & 316  &  316  &  18  &  9  \\
   iii. & 1,000  &  1,000  &  32  &  16  \\
   iv. & 3,162  &  3,162  &  56  &  28  \\
   v. & 10,000  &  10,000  &  100  &  50  \\
   vi. & 31,623  &  31,623  &  178  &  89  \\ 
   \hline
   SNAKE & \multirow{2}{*}{10,000}  & \multirow{2}{*}{10,000}  &  \multirow{2}{*}{100 sets}  &  \multirow{2}{*}{50 sets} \\ 
   competition & & & & \\
\end{tabular}
}
\label{tab:dataset_sizes}
\end{table}

One noticable trend from Figure \ref{fig:expB} is that the accuracies of all MIAs tend to degrade for the larger configurations used. We rule out the quality of the synthetic data as an explanation (see Figure \ref{fig:all_distances} where we show that the data quality actually \textit{improves} when increasing the dataset sizes). So the decreasing accuracies can possibly be explained by the diminishing proportion of targets to training data, ostensibly lowering the impact had on the resulting synthetic data.

\subsection{Experiment C results}

Since the TAPAS attack is our most competitive contender, we strengthen its training conditions, feeding it 4,000 shadow datasets instead of 500 as in the previous experiments. Because of computational costs, we only do this for $|D_{train}| = 1,000$ and $\varepsilon = 10$, on the SNAKE data. Its accuracy improves slightly, but it still does not beat the accuracy of MAMA-MIA, which is still only using 50 shadow runs. These results are shown in Table \ref{tab:expC} along with corresponding runtimes. (The differences in performance when TAPAS uses 500 versus 4,000 shadow datasets can be seen in Figure \ref{fig:tapas_s50}).

\begingroup

\setlength{\tabcolsep}{4pt} 
\begin{table}[ht!]
\caption{\textbf{Experiment C} The TAPAS attack when generating and using 4,000 shadow datasets compared with MAMA-MIA, which uses only 50 shadow runs. $|D_{train}| = 1,000$ and  $\varepsilon = 10$ on SNAKE data. Private-GSD uses its default 20,000 generations.}
\centering
\begin{tabular}[b]{c | c c | c c}
\multicolumn{1}{c}{} &\multicolumn{2}{c}{mean AUC $\pm 1$ std. dev.} &\multicolumn{2}{c}{runtime} \\
    & TAPAS & MAMA-MIA & TAPAS & MAMA-MIA \\ \hline
   MST  & $0.599 \pm .09$  & $0.614 \pm .11$ & $1.78$ \textbf{days}  & $14.0$ \textbf{min} \\
   PrivBayes & $0.565 \pm .09$ & $0.578 \pm .09$ & $6.44$ \textbf{hours} &  $11.2$ \textbf{sec} \\
   Private-GSD  &  $0.498 \pm .11$ & $0.612 \pm .10$ & $8.85$ \textbf{days}  & $6.26$ \textbf{min} \\
   
\end{tabular}
\label{tab:expC}
\end{table}

\endgroup

\subsection{Additional results}\label{sec:expD}

\noindent{\textbf{Extending experiments to the California Housing Dataset}}
We extend our experiments from Section \ref{sec:expA} to the California Housing Dataset, containing only 20,640 records (as opposed to SNAKE's 201,279) and nine continuous features (SNAKE has fifteen categorical). Since all of these SDG algorithms operate over discrete values, we encode the values into twenty \textit{equal-depth} bins. Again, $|D_{train}| = 1,000$ records. The results are shown in Figure \ref{fig:expD_cali}. MAMA-MIA's performance is even more distinguished on this data.


Curiously, it is on the housing data where DOMIAS-using-BNAF seem to perform most apparently above random guessing, and that could be due to BNAF performing better on fundamentally numerical data.

\begin{figure*}
\centering
\includegraphics[width=\textwidth]{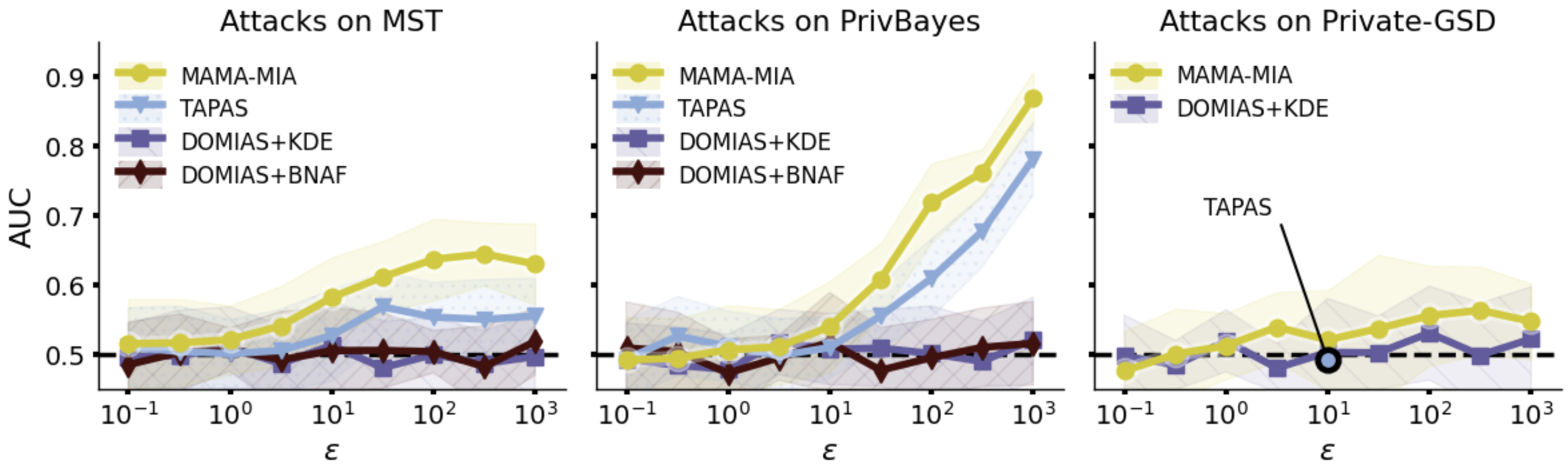}
\caption{A more strenuous setting where $|D_{train}|, |D_{synth}| = 10,000$ records of SNAKE data and $|D_{member}| = 50$. We perform the TAPAS attack on Private-GSD \textit{only for $\varepsilon = 10$} because of the computational costs to generate 500 Private-GSD shadow datasets with 10,000 rows.} 
\label{fig:exp_rigor}
\end{figure*}

\begin{figure*}
\centering
\includegraphics[width=\textwidth]{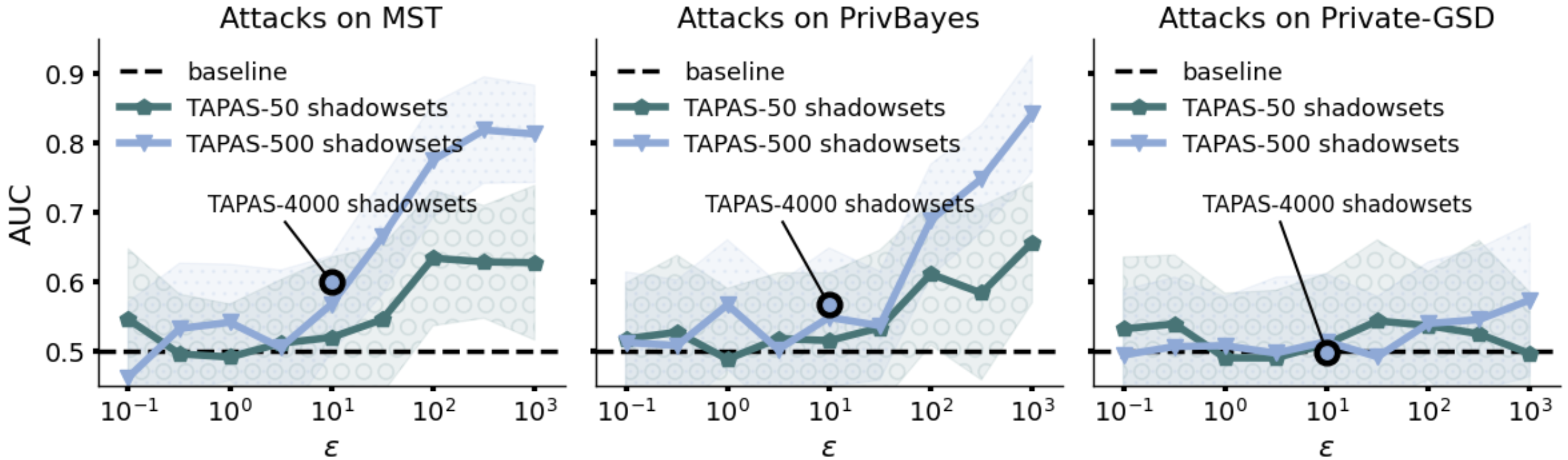}
\caption{The accuracy of the TAPAS membership inference attack on the SNAKE data significantly degrades when using 50 shadow datasets versus 500 or 4000. (The ``TAPAS-500'' scores are the same as those presented in Figure \ref{fig:expA}.)} 
\label{fig:tapas_s50}
\end{figure*}

TAPAS performed unexpectedly worse on the housing data than it did on the demographic SNAKE data. So instead of running TAPAS with 3-way queries, we ran it using all possible 5-way queries, 2-way queries, and 1-way queries, which did not improve its scores. We then tried running the TAPAS pipeline where the continuous data is binned into only five equal-depth bins, which again did not yield better results. In the end, this can be explained by how many more possible queries are used as features to the TAPAS model when attacking the \textit{SNAKE data} (i.e. there are 455 possible 3-way marginals over the 15 features) versus attacking the California housing data, which has only nine features (so there are only 84 possible 3-way marginals). (In the paper in which this attack is proposed, experiments are done on datasets with fifteen and eighteen features \cite{stadler2022synthetic}, and the TAPAS paper uses a dataset with fifteen features \cite{houssiau2022tapas}.) If this were the case, then these results would attest to MAMA-MIA's resilience in such settings where the number of features is limited.

\vspace{7pt}\noindent{\textbf{Exploring a more strenuous setting}} 
We conduct experiment A, though now increasing $|D_{train}|, |D_{synth}|$ from $1,000$ records to $10,000$ records, and decreasing the proportion of targets from 3.2\% to 1\% of $D_{train}$ (see configuration (v) from Table \ref{tab:dataset_sizes}). Made clear in Figure \ref{fig:exp_rigor}, the accuracies are lower for this setting because of the reduced proportion of targets, as explained in section \ref{sec:expB}. However, MAMA-MIA more definitively outperforms the other attacks when the datasets are large. In fact, our PrivBayes attack accuracy barely diminishes at all from the smaller setting. It is unclear why this happens, but is likely due to PrivBayes' use of conditionals as FPs, which have up to four parents, and so hold greater specificity than the 2-way marginals chosen by MST and the 3-way marginals in Private-GSD. (See our extended results on PrivBayes' FP selection in Appendix section \ref{sec:extended_results} for a discussion).

We only conduct the TAPAS attack on Private-GSD for when $\varepsilon = 10$ because of the high cost of generating 500 shadow datasets for each setting. To give a sense, just generating the Private-GSD shadow datasets for one TAPAS attack took \textit{387 days} of processor time. The second part of the attack that processes the shadow datasets took 19 minutes. 

Note that this exorbitant time was in part because we had to increase the number of generations (i.e. iterations) Private-GSD uses from 20,000 by default to 100,000. We explain why below, where we evaluate the qualities of the synthetic datasets for various dataset sizes. That being said, we did manage to generate 500 TAPAS shadow datasets for \textit{all} values of $\varepsilon$ using 20,000 generations. However, none of these TAPAS attacks, nor any other MIA, performed better than random guessing, even when $\varepsilon = 1,000$.

\begin{figure*}[!ht]
\centering
\includegraphics[width=\textwidth]{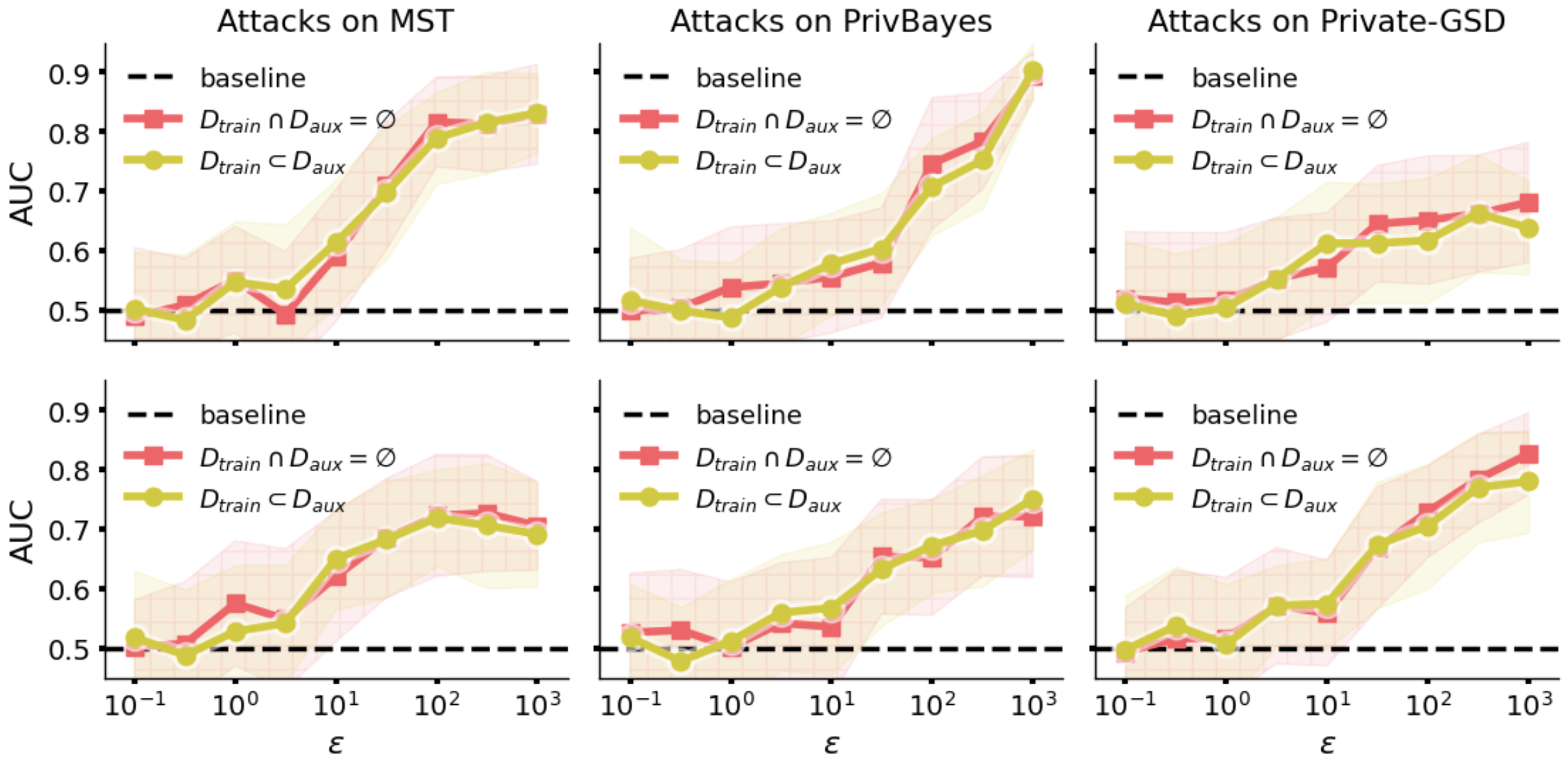}
\caption{Comparing MAMA-MIA scores from experiment A on the \textbf{SNAKE data} (top row), and on the \textbf{housing data} (bottom row) when $D_{train} \subset D_{aux}$ and when $D_{train} \cap D_{aux} = \varnothing$. (The ``$D_{train} \subset D_{aux}$'' scores are the same as those presented in Figures \ref{fig:expA} and \ref{fig:expD_cali} respectively.)} 
\label{fig:nonoverlap_snake}
\end{figure*}

\begin{figure}
\centering
\includegraphics[width=8.8cm]{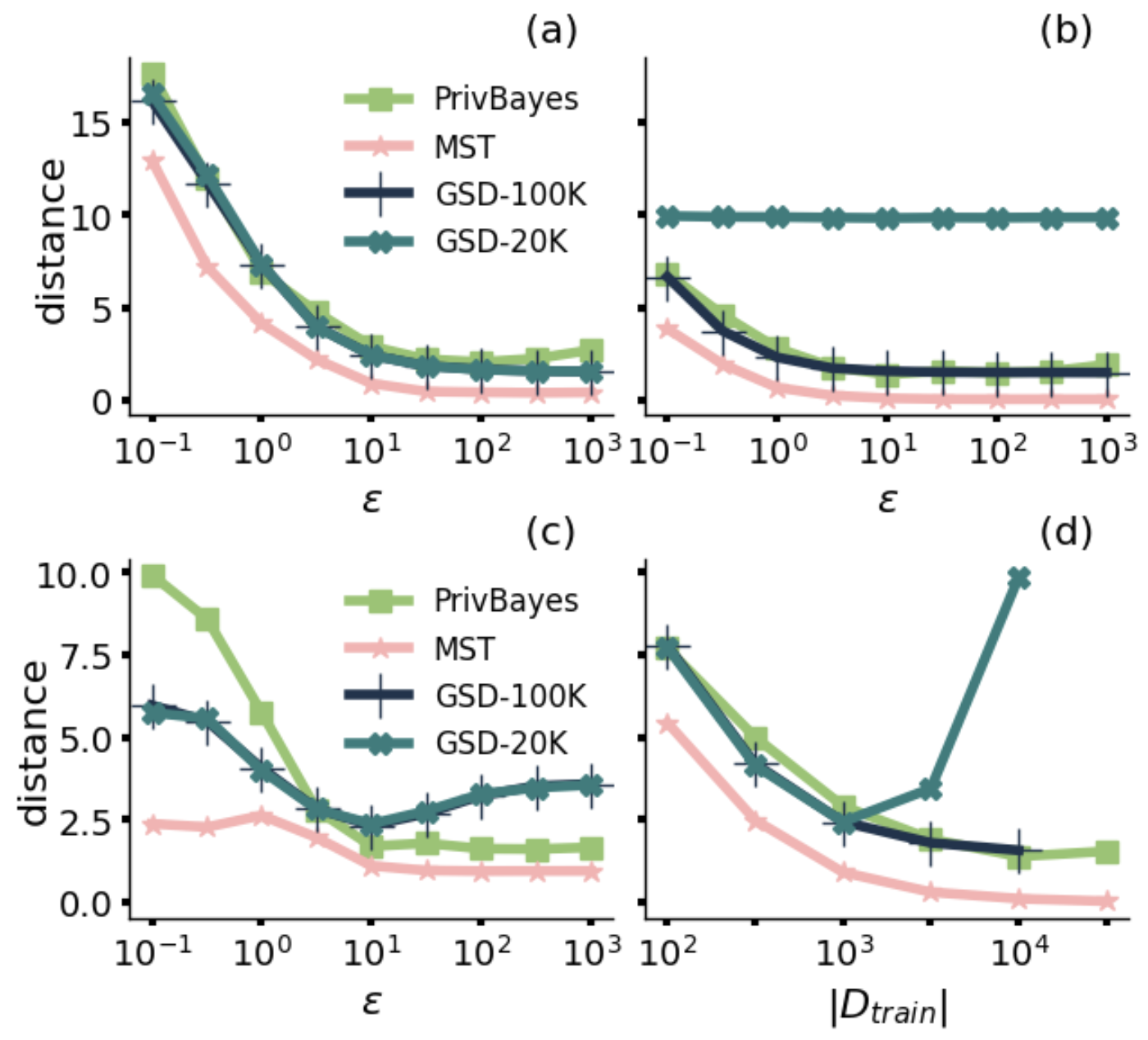}
\caption{Distances of 
\textbf{(a)} SNAKE data in setting (iii) (when $|D_{train}| = 1,000$), 
\textbf{(b)} SNAKE data in setting (v) (when $|D_{train}| = 10,000$), 
\textbf{(c)} housing data in setting (iii) (when $|D_{train}| = 1,000$), and
\textbf{(d)} SNAKE data for various \textit{sizes}, when $\varepsilon = 10$.
Results are shown for Private-GSD using 20,000 generations (``GSD-20K'') and 100,000 generations (``GSD-100K''), both with early stopping. The astute reader will notice the overlap between \textbf{(a)} and \textbf{(d)}, and between \textbf{(b)} and \textbf{(d)}.}
\label{fig:all_distances}
\end{figure}

\vspace{7pt}\noindent{\textbf{Evaluating a much more efficient TAPAS attack}}
Since the computational costs associated with the TAPAS MIA were substantial, perhaps such costs aren't entirely necessary. So instead of training the attack with 500 and 4,000 shadow datasets, we train it with only 50, similar to the 50 shadow modelling runs MAMA-MIA uses. This yields comparable, though still worse, runtimes because MAMA-MIA terminates the SDG process early. This notwithstanding, we find that the TAPAS attack accuracy significantly degrades when using 50 shadow datasets versus 500 (shown in Figure \ref{fig:tapas_s50}), suggesting that high computational costs are inherent to that attack's success.

\vspace{7pt}\noindent{\textbf{When the training data does not overlap with the auxiliary data}} 
We show the results of changing our threat model to when $D_{train} \cap D_{aux} = \varnothing$. As seen in Figure \ref{fig:nonoverlap_snake}, our attack is not noticeably impacted. This is expected for the SNAKE data, since it has 201,279 records, and so separating the training data from it does not impact the shape of $D_{aux}$ greatly. As long as $D_{train}$ is pulled from the same distribution, gleaning information from $D_{synth}$, and comparing it to $D_{aux}$, lends insight into the training data $D_{train}$. (This experiment was done using the configuration in experiment A.)

\vspace{15pt}\noindent{\textbf{Shedding light on synthetic data qualities}}
We carry this out by measuring the Wasserstein distance \cite{ramdas2017wasserstein} of each column of the synthetic dataset to that of the same column on the training dataset, and then we sum all those distances\footnote{Note that related works suggest that quantifying the quality of generative models is inherently application-specific \cite{theis2015note, dankar2022multi, hernandez2022synthetic}. While other works too evaluate the quality of synthetic data using Wasserstein Distance (WD), training machine learning models on both $D_{train}$ and $D_{synth}$, then comparing their predictive success, has become commonplace \cite{hittmeir2019utility, pereira2023assessment}. So has comparing arbitrary $k$-way marginals between the two \cite{mckenna2021winning}. For our purposes, training models would have been excessive for the scope of this work, and was more likely to introduce bias. Our aim is simply to provide a modest, reliable distance metric with which to weigh against empirical privacy loss. Furthermore, we deliberately abstained from measuring marginals in our quality assessment. Because we already know that these SDG algorithms maintain marginal consistency, and because relying fully on marginals is contrary to the argument we are making, we opted to show the relative distances from a different angle. Plus, measuring WD on individual columns renders the calculation deterministic. What's important is that our distance measure yields qualities consistent with the findings of related work \cite{pereira2023assessment, tao2021benchmarking}}. 
We provide these results in Figure \ref{fig:all_distances} to accompany attack accuracy results. Intuitively, an attack can only perform well on synthetic data that has a higher resemblance (thus, a lesser distance) to the training data. Private-GSD and PrivBayes tend to have comparable qualities, while MST produces data with the highest quality. This gives insight into why MAMA-MIA scores slightly better on MST than on data from the other two, generally. 

The non-monotonicity of the qualities of Private-GSD housing data in Figure \ref{fig:all_distances} part \textbf{(C)} is somewhat puzzling, and could be the result of an incompatibility between what our distance measurement captures and what Private-GSD captures on the binned housing data. The far-off distances in Private-GSD's data in part \textbf{(b)} of the figure, and the odd spike in its distance in part \textbf{(d)}, show how the quality of Private-GSD data degrades when only using 20,000 generations (the default amount) when $|D_{train}|$ and $|D_{synth}| \geq 3,000$ records. However, when 100,000 generations are used, the quality is as good as data produced by PrivBayes. We always allow for early stopping of Private-GSD, which can explain why, when $|D_{train}|$ and $|D_{synth}| \leq 1,000$ records, using 20,000 generations versus 100,000 generations yields data with the same distances. 

So because of this, in all experiments except for experiment C, we set Private-GSD to use 100,000 generations.

\vspace{7pt}\noindent{\textbf{SNAKE Challenge results}}
Lastly, in Table \ref{tab:snake_scores_priv} we show the outcome of the 2023 SNAKE Challenge, where MAMA-MIA won first place for all privacy levels on MST and PrivBayes.\footnote{PATE-GAN \cite{jordon2018pate} was the third SDG algorithm attacked in this competition, but we omit the results here since it is not a marginals-based method, and because no teams' attack scored significantly above random guessing. This is corroborated by \cite{ganev2024elusive}, where the authors were not able to generate high quality data with existing implementations of PATE-GAN.} Dataset sizes were as listed in the ``SNAKE competition'' configuration in Table \ref{tab:dataset_sizes}. Instead of being evaluated using AUC, results were scored using \textit{membership advantage} (MA) \cite{yeom2018privacy}, 
\begin{equation}
    \textit{MA} = (\textit{tpr} - \textit{fpr} + 1) / 2
\end{equation}
where $\textit{tpr}$, $\textit{fpr}$ are the true positive rate and false positive rate, computed by weighting individual predictions by their distance from a $0.5$ threshold: $2 \cdot |0.5 - p_i|, p_i \in P$. Like with AUC, an MA score of $0.5$ is as good as random guessing. 
In all of our experiments, MA almost always ended up being within $\pm$ 0.02 of what the AUC score was. Our accuracies are noticably better here than for the above experiments because the objective of this challenge was ``set MI'' (where targets are grouped into sets of five or more records, and either the \textit{entire set} of records is in $D_{member}$ or none are). We provide more set MI results in Appendix \ref{sec:extended_results}.

\begin{table*}
\caption{Podium results for the SNAKE Challenge, evaluated using MA.}
\centering
\begin{tabular}[b]{ c | cccc | cccc }
\multicolumn{1}{c}{} &\multicolumn{4}{c}{Attacks on MST} &\multicolumn{4}{c}{Attacks on PrivBayes} \\
    $\varepsilon$    &  1  &  10  &  100  &  1000 &  1  &  10  &  100  &  1000\\
    \hline 
    \textbf{MAMA-MIA}  & \textbf{0.61} & \textbf{0.79} & \textbf{0.76} & \textbf{0.81} & \textbf{0.65} & \textbf{0.71} & \textbf{0.83} & \textbf{0.94} \\
    ($2^{nd}$ place) &  0.60  &  0.60  &  0.55  &  0.69 &  0.53  &  0.62  &  0.69  &  0.51  \\
    ($3^{rd}$ place) &  0.60  &  0.53  &  0.56  &  0.54 &  0.57  &  0.59  &  0.61  &  0.55   \\
\end{tabular}
\label{tab:snake_scores_priv}
\end{table*}

%
%

\newpage
\section{Discussion and broader impact}
Our attack shows remarkable improvements over the other leading MIAs on marginals-based synthetic data in terms of efficiency and accuracy. In short, the success achieved by MAMA-MIA is attributed to leveraging simple knowledge of each generator, thereby constructing a novel density estimation $\zeta$ of the training data. This construction is based on measuring the marginals that we can confidently predict were used by the generator. Our approach is generalizable to several marginals-based algorithms, and can be adapted to others still.

This paper raises several concerns about synthetic data generation––some new, some reinforcing prior research. For one, it empirically shows how much stronger an adversary is when the SDG algorithm used to generate a dataset is published, and so allows an SDG operator to understand the consequences in doing so. Likewise, synthetic data preserving marginals or other related statistics, though they can produce high quality data, are susceptible to an attack following our paradigm. What has not been shown before, to the best of our knowledge, is how computationally light-weight attacks on this type of data can be. We re-emphasize here the relevance of an attack's feasibility to policy decisions \cite{GDPR2016a}.

Even in applying DP noise, when $\varepsilon \geq 10$, our density estimation provides a meaningful approximation of the training data, and exposes individual-level information. Choosing $\varepsilon$ is certainly application-specific--dependent on a multitude of factors such as the desired utility and characteristics of the dataset. Still, $\varepsilon \approx 10$ is generally considered reasonable \cite{nistdifferentialprivacy}. (We don't expect $\varepsilon = 1000$ to be used in practice, and explore it here as a way of pronouncing the relative efficacies of different MIAs, and exposing the vulnerabilities of synthetic data when (almost) no differential privacy protection is used.)


We are excited to continue pursuing this topic further. This includes designing new density functions to exploit traits of other state-of-the-art synthetic generators, like MWEM-PGM, or adapting it to GAN-based algorithms. It also includes conducting further analysis of $\varepsilon$'s effect on both the privacy and utility of SDG algorithms, and how our attack's success on each algorithm calls for a reevaluation of that trade-off. And most importantly, future work includes developing new synthetic data generation techniques that are resistant to membership inference attacks that use this approach.

\vfill\eject

\section*{Acknowledgments}
Steven Golob and Sikha Pentyala are Carwein-Andrews Distinguished Fellows.
Steven Golob is supported by an NSF CSGrad4US fellowship. Sikha Pentyala is supported by a JP Morgan Chase PhD fellowship.  This research was, in part, funded by the UW Global Innovation Fund, the UWT Founders Endowment and the National Institutes of Health (NIH) Agreement No.1OT2OD032581.

We would like to sincerely thank the organizers of the SNAKE challenge for posing this very interesting problem. Our research group, which has traditionally focused on privacy defenses, learned greatly from attempting to audit them.

%
%

\vspace{10pt}
\bibliography{main}
\bibliographystyle{plain}

%
%

\vspace{10pt}
\appendix
\section{Appendix}

\subsection{Discretionary choices}

\noindent{\textbf{An alternative $\zeta$}}\label{sec:alternative_zeta}
Various aggregations based on measuring the predicted focal-points in $D_{synth}$ and $D_{aux}$ could be used for $\zeta$. One such aggregation that could replace line 5 of Algorithm \ref{alg:customMIA} is the mathematically-derived


\begin{equation*}
    g_t = \sum_{\mathbf{F} \in \mathcal{F}}^{} 
    \left\{
    	\begin{array}{ll}
    		\log (\hat{P}_{D_{synth}}(\mathbf{F}(t)) / \hat{P}_{D_{aux}}(\mathbf{F}(t))) & \mbox{if } w_{\mathbf{F}} \geq \omega \\
    		0 & \mbox{if } w_{\mathbf{F}} < \omega
    	\end{array}
    \right.
\end{equation*}
\begin{equation}
    \zeta[t] = e^{g_t}
\end{equation}
where $\omega$ is some threshold for how many times the focal-point $\mathbf{F}$ was chosen out of 50 runs ($\omega = 40$, for instance). However, we chose the aggregation in Algorithm \ref{alg:customMIA} because the resulting density estimation achieved slightly better attack accuracy throughout.

\begin{figure*}[!ht]
\centering
\includegraphics[width=\textwidth]{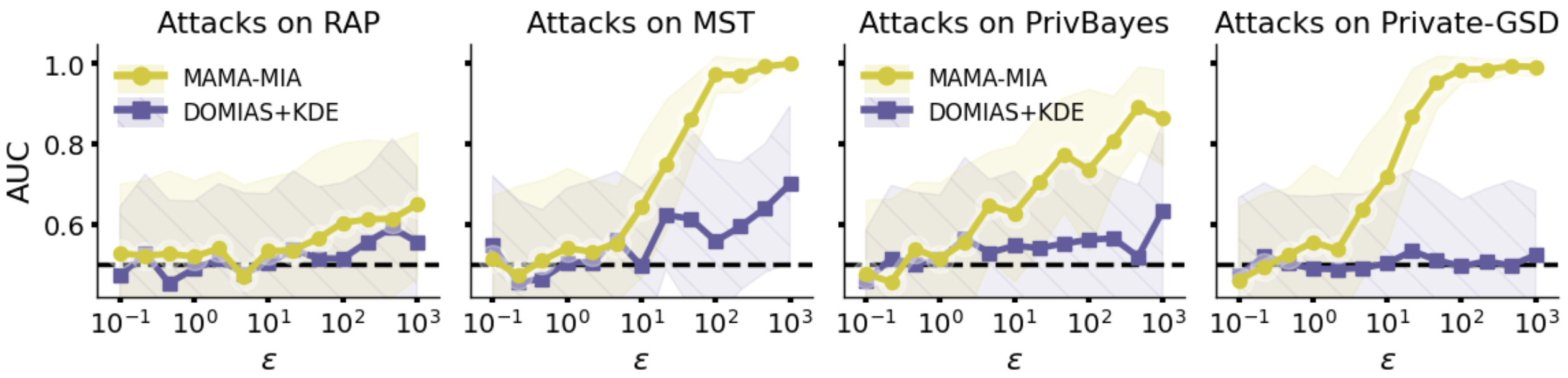}
\caption{Attack accuracies of MAMA-MIA and DOMIAS-using-KDE on RAP, as well as MST, PrivBayes, and Private-GSD. Set MI is performed on the SNAKE data, where $|D_{train}| = 100$ (setting (i) in Table \ref{tab:dataset_sizes}).} 
\label{fig:rap_results}
\end{figure*}

\newpage
\vspace{7pt}\noindent{\textbf{Activation function}}\label{sec:activation}
Once we compute $\zeta$ for the set of targets $D_{target}$, we convert them to a probability of membership in $D_{train}$, $P(\zeta) \in [0, 1]$. We achieve this by designing an activation function,
\begin{equation}    
    P(\zeta) = \frac{1}{(1 + e^{-c (\log \zeta - m)})}
\end{equation}
which maintains the monotonicity of $\zeta$. This is simply a modified sigmoid function, where $c$ is a confidence parameter, defining how far away from probability $0.5$ we want our predictions.

This function also maintains that $P(\zeta) = 0.5$ when the densities for $D_{synth}$ and $D_{aux}$ are the same for a given target, which is consistent with the intuition behind how $\Lambda$ is defined in DOMIAS \cite{van2023membership}. In Section \ref{sec:method}. $m$ is the value of $\zeta$ for a certain percentile among all of the targets. This percentile is the expected proportion of targets that are actually members. For example, since in all of our experiments exactly half of the targets were members, we set $m$ in $P(\zeta)$ to the median density estimation of the targets. This maps half of the targets to $P > 0.5$, and half of the targets to $P < 0.5$ (except for the median, which is predicted at $P = 0.5$). 

Or, in the more realistic setting, when an adversary has no knowledge or expectation of how many targets are members, setting
$P(\zeta) =$ min$(\sqrt[c]{\zeta}/2$, 1) can be a useful mapping of $\zeta$ to probabilities, with $c$ similarly acting as a confidence level. This function maintains that when $\zeta > 1$, probability of membership is $> 50\%$, which is still consistent with the intuition behind $\Lambda$.

DOMIAS and other works on MIAs stop short of defining an activation function, and score their results using AUC. We define ours here because of the good results it achieved in the SNAKE Challenge, and because our density estimation function $\zeta$ is a direct replacement for DOMIAS.

\begin{figure}[]
\centering
\includegraphics[width=8.8cm]{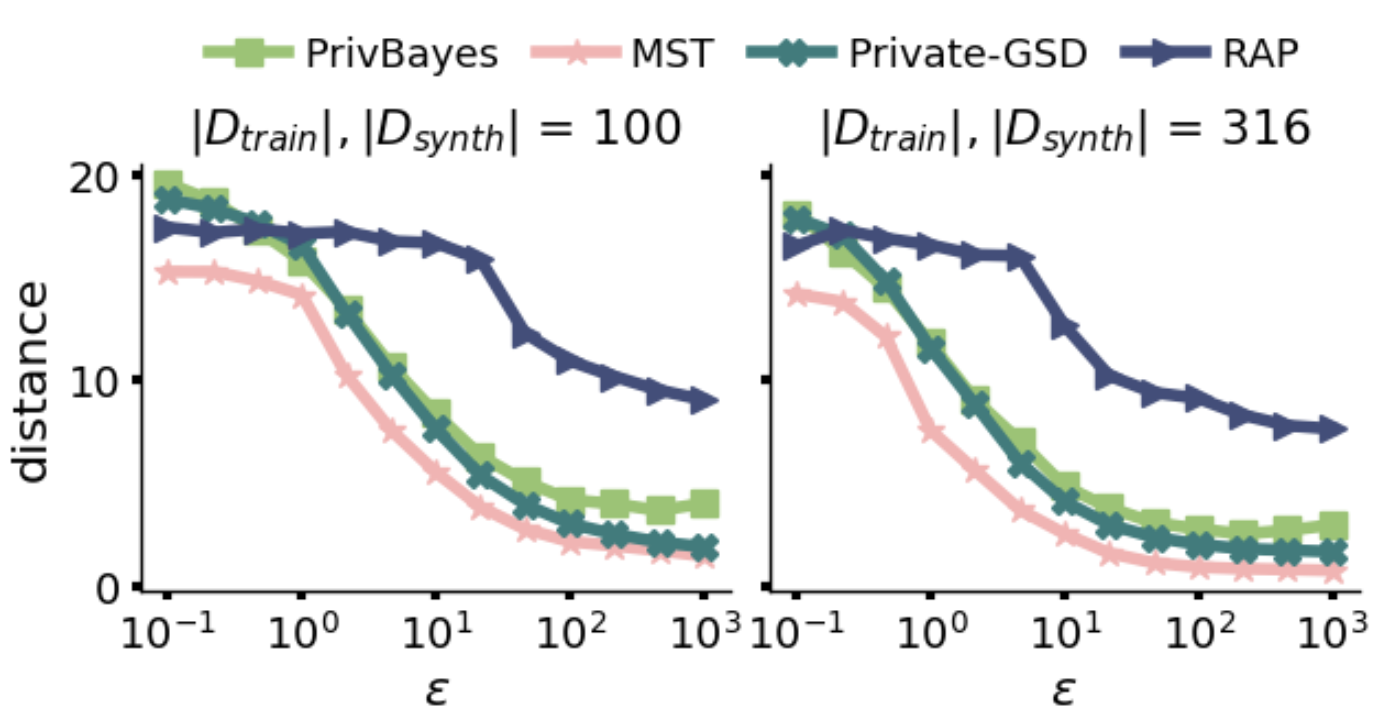}
\caption{Distance from $D_{synth}$ generated by each SDG algorithm to $D_{train}$. The datasets have 100 rows (left) and 316 rows (right) of SNAKE data. We were not able to gather results for RAP generating larger dataset sizes due to computational limitations.} 
\label{fig:distances_wRAP}
\end{figure}

\subsection{MAMA-MIA on RAP}\label{sec:RAP}

We initially included a fourth SDG algorithm in our investigation that also makes use of marginals and provides differential privacy, RAP (Relaxed Adaptive Projection) \cite{aydore2021differentially}. 

However, we do not include it in the main text because we found that this SDG algorithm did not produce high quality data, both for the SNAKE dataset and the housing dataset. See, for example, our distance measurement (described in Section \ref{sec:expD}) of RAP's $D_{synth}$ to its training data in Figure \ref{fig:distances_wRAP}, compared with those generated by the other SDG algorithms. Despite our best efforts to bolster the quality of the RAP data through many training parameters (of course beginning with its defaults), we could not improve its quality much. This result is consistent with other scholarship ranking the qualities of MST, PrivBayes, and RAP \cite{tao2021benchmarking}. If our simple measurement yields such a great divergence between the real data and RAP's synthetic data, it is certainly unlikely that RAP could maintain other important properties, like marginals, between the real and the synthetic data. 

The best result MAMA-MIA achieved on RAP data was when $D_{train}$ had only 100 rows of SNAKE data, and when performing set MI on it, shown in Figure \ref{fig:rap_results}. Notice how, in this more susceptible setting, our attack on other SDG algorithms can get much closer to perfect accuracy. 

This was also the only setting where DOMIAS, using KDE, performed noticeably better than random guessing on MST and PrivBayes. For these investigations in Figures \ref{fig:distances_wRAP} \& \ref{fig:rap_results}, we include thirteen values of $\varepsilon \in \{10^{i/3} | -3 \leq i \leq 9\}$, but do not have results for the TAPAS or DOMIAS-using-BNAF attacks due to computational limitations.

\begin{table*}
\caption{Runtimes of MAMA-MIA on SNAKE data when $\varepsilon = 10$ (i.e. experiment B), separated into the shadow modelling stage, where we select FPs, and the inference stage, where we use them to compute $\zeta$. The runtime of the select FP stage is for 50 shadow modelling runs. The runtime of the compute $\zeta$ stage given is the average runtime for one trial, attacking all targets in $D_{target}$. Results for Private-GSD on 31,623 records are omitted because of computational limitations.}
\centering
\begin{tabular}[b]{c | c c | c c | c c}
\multicolumn{1}{c}{} &\multicolumn{2}{c}{MAMA-MIA on MST} &\multicolumn{2}{c}{MAMA-MIA on PrivBayes} &\multicolumn{2}{c}{MAMA-MIA on Private-GSD} \\
   $|D_{train}|$ & select FPs & compute $\zeta$ & select FPs & compute $\zeta$ & select FPs & compute $\zeta$ \\ \hline
    100 & 13.2 \textbf{min} & 2.57 \textbf{sec} & 2.11 \textbf{sec} & 6.73 \textbf{sec} & 2.57 \textbf{min} & 16.3 \textbf{sec}  \\ 
    316 & 13.3 \textbf{min} & 3.85 \textbf{sec} & 2.63 \textbf{sec} & 5.33 \textbf{sec} & 2.69 \textbf{min} & 38.8 \textbf{sec}  \\ 
    1,000 & 14.0 \textbf{min} & 4.14 \textbf{sec} & 6.46 \textbf{sec} & 4.74 \textbf{sec} & 4.93 \textbf{min} & 79.9 \textbf{sec}  \\ 
    3,162 & 14.3 \textbf{min} & 3.75 \textbf{sec} & 15.7 \textbf{sec} & 5.22 \textbf{sec} & 8.09 \textbf{min} & 2.17 \textbf{min}  \\ 
    10,000 & 14.5 \textbf{min} & 3.02 \textbf{sec} & 50.0 \textbf{sec} & 4.37 \textbf{sec} & 10.2 \textbf{min} & 2.52 \textbf{min}  \\ 
    31,623 & 41.7 \textbf{min} & 1.01 \textbf{sec} & 4.37 \textbf{min} & 4.43 \textbf{sec} & N/a & N/a  \\
\end{tabular}
\label{tab:runtime_separated}
\end{table*}

We now describe how RAP works, how we tailor MAMA-MIA to attacking it, and \textit{why} its data quality was lower.

Similar to Private-GSD, RAP does not build a graphical model to estimate the joint probability distribution. Instead, it encodes $D_{train}$'s features into binary form, then initializes an arbitrary dataset $D'$ of the same dimension, which is then updated repeatedly until the focal-point measurements from $D'$ resemble those taken on $D_{train}$. 

RAP's focal-points (FPs) are called ``queries'', which are simply $k$-way marginals on the binarized features, where $k = 3$ by default. Over several iterations, it measures new queries, and re-updates $D'$ using differentiable learning with respect to the errors of those queries. This update happens using \textit{Sparsemax} (a variant of softmax) to achieve gradient descent \cite{martins2016softmax}. Once RAP is finished updating $D'$, it is projected from a table of floating point values into the binarized domain. This one-hot encoding is then decoded into $D_{synth}$.

The queries that yield the greatest differences between $D_{train}$ and $D'$ are favored by RAP, in an effort to reduce the maximum error across all queries. The default number $q$ of queries is $50$. Note that $q = 50$ is quite small, relative to the $l-1$ marginals measured by MST, since there are \textit{far} more possible queries over the one-hot encoded domain. Consequentially, the amount of information contained in a marginal on binary features is much less. 

So RAP expects the scientist to specify a ``workload''--that is, hand-select a subset of features to be considered in the query selection process--using domain- and task-specific knowledge. This greatly narrows the amount of possible focal-points considered, and so poses a challenge for us during shadow-modelling; since knowledge of the workload is not included in our threat model, and since we only have access to default information, the FPs we observe will likely be wildly different from those actually measured on $D_{train}$. Even when no workload is chosen by the scientist, and default behavior is used to train the generator (as is done in our experiments), the shear amount of FPs considered causes the fitting of $D_{synth}$ to be highly volatile, even when $\varepsilon$ is large. This notwithstanding, our density estimation $\zeta$ of RAP synthetic data uses the frequencies of queries chosen during shadow modelling, and we use those as the weighted focal-points in Algorithm \ref{alg:customMIA}.


\subsection{Experimental parameters}\label{hyperparameters}





For each attack, shadow modelling consisted of 50 runs. For each membership experiment described in this paper, results were averaged over 30 runs. All values of $\varepsilon$ and dataset sizes tried are evenly spaced on the logarithmic scale, while the proportion of $D_{target}$ to $D_{train}$ changes from $10\%$ of the training to $0.56\%$, to stage different inference challenges. (We omit experiments on the California Housing Dataset using the largest configuration because it only contains 20,640 records.)

Since MST, PrivBayes, Private-GSD, and RAP operate on discretely valued data, we segment continuous values into 20 equal-depth bins. Since the KDE estimator used by DOMIAS operates on numerical values, we encode the categorical features of the SNAKE data ordinally.

During set MI experiments, sets used in $D_{target}$ are randomly sampled from those that contain at least five records.

\begin{table*}[!ht]
\caption{Average AUC scores ($\pm$ one standard deviation) over 30 trials for MAMA-MIA and TAPAS attacks on the \textbf{SNAKE data}. $D_{train} = 1,000$, and TAPAS uses 500 shadow datasets.}
\centering
\begin{tabular}[b]{c | cc | cc | cc}
     & \multicolumn{2}{c}{Attacks on MST} & \multicolumn{2}{c}{Attacks on PrivBayes} & \multicolumn{2}{c}{{Attacks on Private-GSD}} \\
   $\varepsilon$ & TAPAS & MAMA-MIA & TAPAS & MAMA-MIA & TAPAS & MAMA-MIA \\ \hline
0.1  & $0.461 \pm .11$ & $0.502 \pm .10$ & $0.512 \pm .10$ & $0.516 \pm .12$ & $0.483 \pm .10$ & $0.503 \pm .10$ \\
0.32  & $0.532 \pm .09$ & $0.483 \pm .11$ & $0.507 \pm .09$ & $0.499 \pm .09$ & $0.494 \pm .11$ & $0.476 \pm .11$ \\
1.0  & $0.541 \pm .08$ & $0.547 \pm .10$ & $0.566 \pm .09$ & $0.487 \pm .09$ & $0.533 \pm .07$ & $0.525 \pm .08$ \\
3.16  & $0.504 \pm .11$ & $0.535 \pm .11$ & $0.502 \pm .09$ & $0.539 \pm .10$ & $0.489 \pm .10$ & $0.551 \pm .10$ \\
10.0  & $0.567 \pm .07$ & $0.614 \pm .11$ & $0.548 \pm .10$ & $0.578 \pm .09$ & $0.515 \pm .15$ & $0.600 \pm .08$ \\
31.62  & $0.664 \pm .08$ & $0.698 \pm .11$ & $0.536 \pm .08$ & $0.603 \pm .09$ & $0.541 \pm .09$ & $0.610 \pm .09$ \\
100.0  & $0.774 \pm .08$ & $0.789 \pm .08$ & $0.688 \pm .08$ & $0.708 \pm .08$ & $0.542 \pm .12$ & $0.634 \pm .10$ \\
316.23  & $0.818 \pm .08$ & $0.815 \pm .09$ & $0.748 \pm .08$ & $0.752 \pm .08$ & $0.549 \pm .07$ & $0.679 \pm .07$ \\
1000.0  & $0.812 \pm .07$ & $0.832 \pm .07$ & $0.841 \pm .08$ & $0.903 \pm .05$ & $0.550 \pm .10$ & $0.664 \pm .09$ \\
\end{tabular}
\label{tab:auc_snake}
\end{table*}

\begin{table*}[!ht]
\caption{Average AUC scores ($\pm$ one standard deviation) over 30 trials for MAMA-MIA and TAPAS attacks on the \textbf{housing data}. $D_{train} = 1,000$, and TAPAS uses 500 shadow datasets.}
\centering
\begin{tabular}[b]{c | cc | cc | cc}
     & \multicolumn{2}{c}{Attacks on MST} & \multicolumn{2}{c}{Attacks on PrivBayes} & \multicolumn{2}{c}{{Attacks on Private-GSD}} \\
   $\varepsilon$ & TAPAS & MAMA-MIA & TAPAS & MAMA-MIA & TAPAS & MAMA-MIA \\ \hline
0.1  & $0.504 \pm .11$ & $0.517 \pm .11$ & $0.517 \pm .12$ & $0.518 \pm .09$ & $0.491 \pm .10$ & $0.513 \pm .09$ \\
0.32  & $0.513 \pm .10$ & $0.488 \pm .11$ & $0.480 \pm .12$ & $0.478 \pm .09$ & $0.496 \pm .08$ & $0.553 \pm .12$ \\
1.0  & $0.509 \pm .09$ & $0.528 \pm .11$ & $0.480 \pm .09$ & $0.511 \pm .10$ & $0.499 \pm .10$ & $0.502 \pm .11$ \\
3.16  & $0.477 \pm .13$ & $0.542 \pm .10$ & $0.488 \pm .11$ & $0.559 \pm .10$ & $0.500 \pm .10$ & $0.541 \pm .09$ \\
10.0  & $0.494 \pm .08$ & $0.650 \pm .09$ & $0.495 \pm .11$ & $0.567 \pm .11$ & $0.535 \pm .11$ & $0.586 \pm .06$ \\
31.62  & $0.500 \pm .09$ & $0.682 \pm .10$ & $0.501 \pm .08$ & $0.633 \pm .10$ & $0.503 \pm .09$ & $0.646 \pm .08$ \\
100.0  & $0.498 \pm .11$ & $0.719 \pm .08$ & $0.535 \pm .11$ & $0.672 \pm .08$ & $0.504 \pm .11$ & $0.710 \pm .09$ \\
316.23  & $0.514 \pm .13$ & $0.706 \pm .11$ & $0.500 \pm .09$ & $0.697 \pm .09$ & $0.528 \pm .10$ & $0.794 \pm .09$ \\
1000.0  & $0.522 \pm .10$ & $0.691 \pm .09$ & $0.522 \pm .10$ & $0.751 \pm .08$ & $0.528 \pm .11$ & $0.803 \pm .07$ \\
\end{tabular}
\label{tab:auc_cali}
\end{table*}

The results on RAP from Figure \ref{fig:rap_results} deviate from the default parameters, in that, instead of selecting 50 queries per epoch, we select 70 queries per epoch, and increase the maximum possible updates per epoch to 2600 from 1000. We do this as a counter measure to induce fitting on the chosen focal-points, selected from a much larger domain of queries (773,239 possible for the one-hot encoded SNAKE data), rather than fitting to a specified workload as RAP intends. Otherwise, all default values for MST, PrivBayes, Private-GSD, and RAP are used in our experiments.

For MST and PrivBayes, we use the implementations given in the Reprosyn library.\footnote{\url{https://github.com/alan-turing-institute/reprosyn}}. 
For Private-GSD and RAP, we use the authors' implementations.\footnote{\url{https://github.com/giusevtr/private_gsd}}$^{,}$\footnote{\url{https://github.com/amazon-science/relaxed-adaptive-projection}}
We use the existing implementations of these DP SDGs. The main idea is to audit the existing algorithms and implementations\footnote{We are aware of the bugs and incorrect implementations such as in PrivBayes as mentioned in \cite{stadler2022synthetic}.}, and the assumptions that the meta data of the training set is available to the adversary.

For all experiments (except in experiment C because of computational limitations), we increase the default number of generations Private-GSD uses from 20,000 with early stopping to 100,000 with early stopping. As shown in Figure \ref{fig:all_distances}, this only makes a difference on the data quality for when $D_{train}$ and $D_{synth}$ have more than 1,000 rows. Therefore, this parameter change did not impact the accuracy of MAMA-MIA in experiment C, whether we used 100,000 or 20,000 generations with early stopping.

\subsection{Complexity and runtime}
\label{sec:efficiency}

The computational costs of our approach can be broken down into two stages. The first stage is where we conduct shadow modelling to determine which focal-points to use in our density estimation. We omit an analysis of the complexity of the SDG algorithms themselves, but can think of them each as some function of the size of the sample training data $\mathcal{O}(f(|\hat{D}_{train}|))$. 
Therefore, if we simulate the algorithm $u$ times, then the complexity of this stage is simply $\mathcal{O}(u \cdot f(|\hat{D}_{train}|))$, since the only computations we add are constant-time steps of recording which focal-points were chosen.

Recall that we terminate the SDG process early once the focal-points are recorded.

The second stage is where we perform the density estimation to calculate membership predictions for $D_{target}$. For each focal-point in MST and Private-GSD, we construct a marginal probability table for $D_{aux}$, $D_{synth}$. We make the reasonable assumption of the ability to use amortized constant-time hash tables during this construction, and so constructing these takes linear time with respect to the size of the datasets, at most size $n$. 
Then for each record in $D_{target}$, we look up the probability of its value. Together, these steps for one focal-point amount to $\mathcal{O}(|D_{aux}| + |D_{synth}| + 2|D_{target}|)$ time, which is just $\mathcal{O}(n)$, since $D_{target} \subseteq D_{aux}$, and assuming $|D_{synth}| < |D_{aux}|$. This is also the case for PrivBayes, where we build conditional probability tables instead. Like marginal tables, constructing conditional tables with hash tables requires looking at each feature in the focal-point for each record once, since the number of features in each conditional is practically small. So the total runtime of calculating $\zeta$ is $\mathcal{O}(|\mathcal{F}|\cdot n)$ where $|\mathcal{F}|$ is the amount of focal-points chosen by the SDG algorithm during shadow modelling.


Runtimes are given in Table \ref{tab:runtime_separated} for this ``compute $\zeta$'' stage separately. PrivBayes takes slightly longer than MST, which is likely due to the fact that it chooses conditionals as focal-points, which offer higher specificity than MST's marginals proper, and so presumably stress the hashing capability. Computing $\zeta$ in Private-GSD appears to take much longer than on the other two SDG algorithms. This can be explained by $|\mathcal{F}|$ being much larger in Private-GSD, which selects thousands of focal-points over one-hot encoded data, as opposed to the dozen or so FPs on un-encoded data chosen by MST and PrivBayes.
The reason why the time to compute $\zeta$ is relatively constant, despite the growing dataset size, is because the $\zeta$ also performs computations over the entire $D_{aux}$, which is much larger (i.e. over 200,000 rows) than any $D_{synth}$ tried, and so dominates that portion of the runtime.

\begin{figure*}[!ht]
\centering
\includegraphics[width=\textwidth]{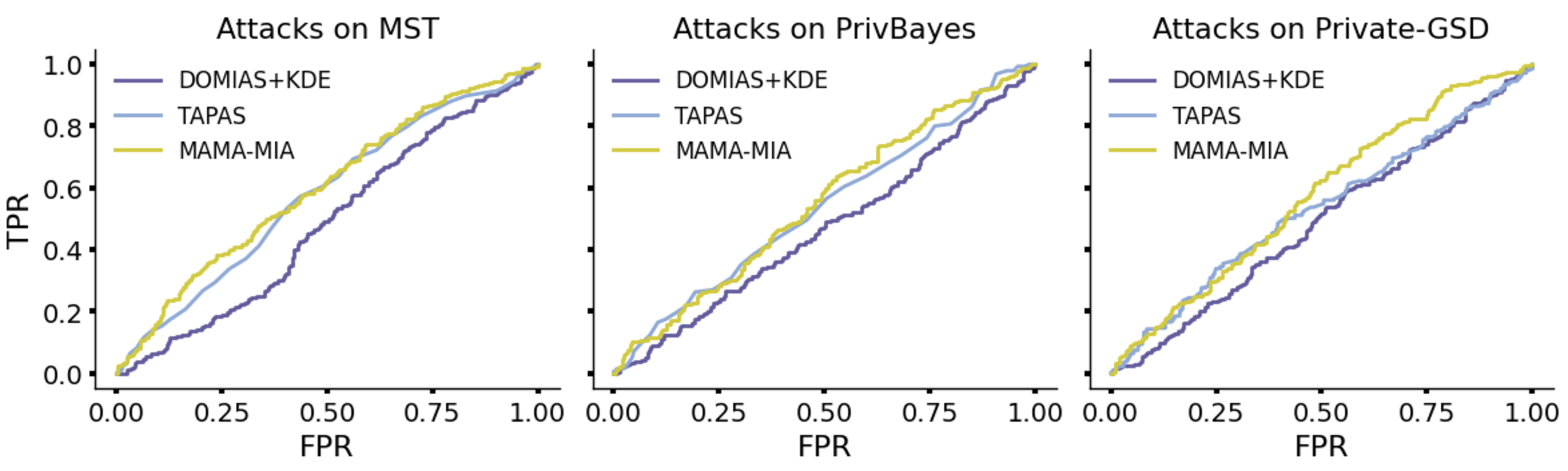}
\caption{ROC curves of MIAs in experiment A when $\varepsilon = 10$.}
\label{fig:roc_curves}
\end{figure*}

\begin{figure}[]
\centering
\includegraphics[width=8.8cm]{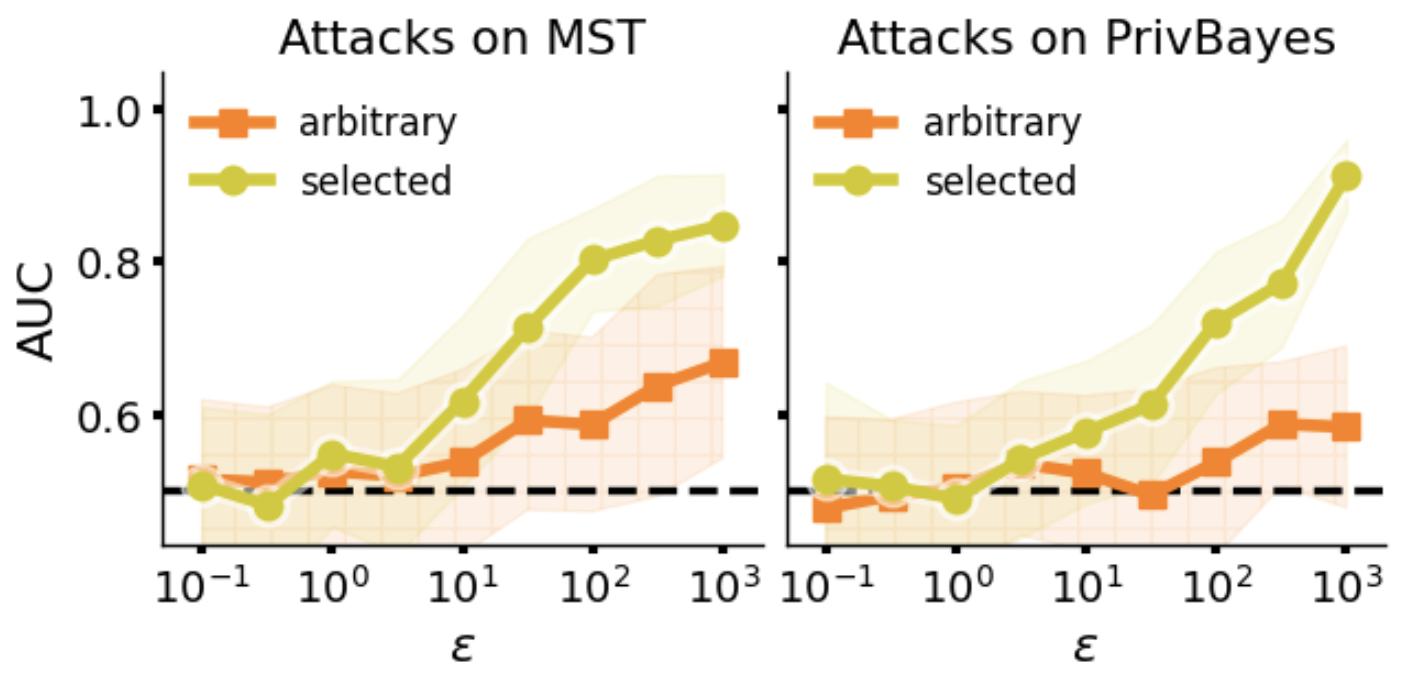}
\caption{Accuracy results of MAMA-MIA when $|D_{train}| = $1,000. The higher curves in each graph show the result of our attack when using FPs obtained during shadow modelling in our density estimation $\zeta$. The lower curves show the degraded performance when arbitrarily-chosen FPs are used in $\zeta$.} 
\label{fig:arbitrary_fp}
\end{figure}

\subsection{Extended results}\label{sec:extended_results}

\noindent{\textbf{Main results in greater detail}} 
We offer AUC results in Tables \ref{tab:auc_snake}, and \ref{tab:auc_cali} for experiment A, conducted on the SNAKE data and the housing data. These tables include MAMA-MIA and TAPAS accuracies, averaged over 30 trials with standard deviations. These correspond to Figures \ref{fig:expA} and \ref{fig:expD_cali} respectively. The runtimes of MAMA-MIA for this experiment, now separated into its shadow modelling and its density estimation step, are given in Table \ref{tab:runtime_separated}.

Additionally, we include full ROC curves in Figure \ref{fig:roc_curves} of the TAPAS attack, DOMIAS, and MAMA-MIA for experiment A, when $\varepsilon = 10$. To average across all trials, we simply compute the curve using all 960 target predictions (across the 30 trials, with 32 targets per trial). These results show in more detail the accuracies of MAMA-MIA predictions at different false positive rates.

\vspace{7pt}\noindent{\textbf{Showing the impact of shadow modelling in MAMA-MIA}}
Shadow modelling to select focal-points (FPs) is a time consuming step in MAMA-MIA, especially when attacking synthetic data generated by SDGs that are in themselves computationally expensive to train (see Table \ref{tab:runtime_separated}). One may wonder what the effect is of this careful selection of focal-points on the accuracy of the attack, i.e.~whether it is worth the computational effort.

We show in Figure \ref{fig:arbitrary_fp} the accuracy of our MAMA-MIA attack as described in Section \ref{sec:method} (using shadow modelling to select focal-points that we predict were used by the generator), and for when we do not do this step, and instead choose the same amount of focal-points \textit{arbitrarily}. The accuracy of this latter configuration is still notable, but is substantially improved by our novel use of shadow modelling.

\begin{figure}
\centering
\includegraphics[width=8.8cm]{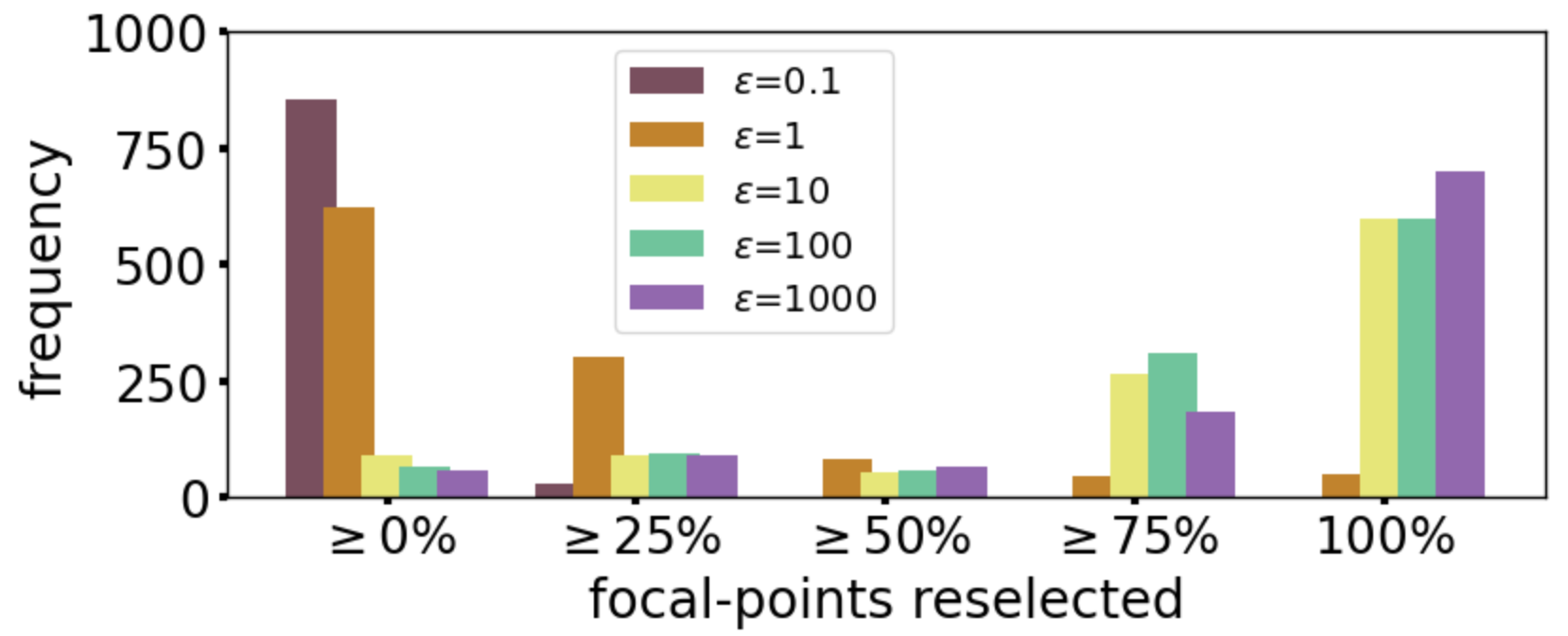}
\caption{Amount of marginals that were chosen by \textbf{MST} at least $x$ percentage of the time, with the tick `100' representing the amount of FPs chosen 100\% of the time during shadow modelling. This chart combines FP counts observed using the California and SNAKE data.} 
\label{fig:marginal_variability}
\end{figure}

\begin{figure}[!ht]
\centering
\includegraphics[width=8.8cm]{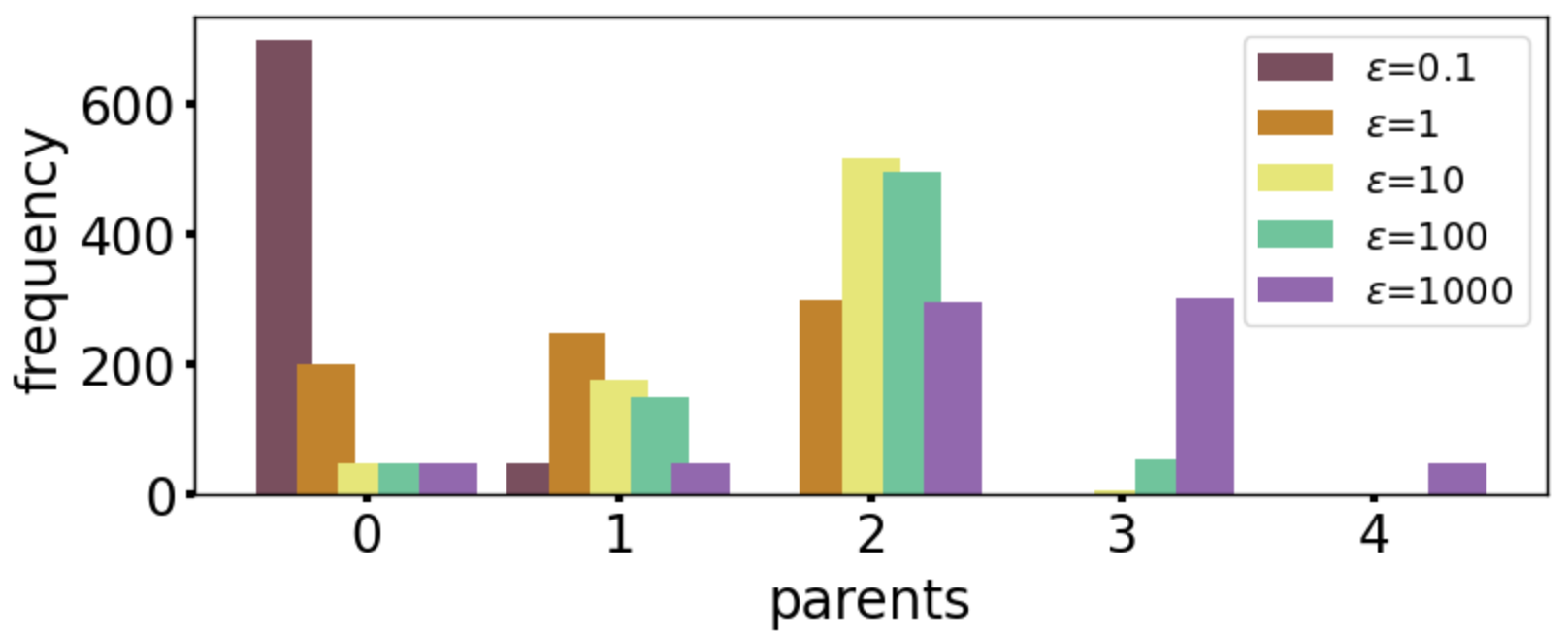}
\caption{Frequencies at which parent sizes were chosen for \textbf{PrivBayes} conditionals, combining shadow modelling on the SNAKE and housing data.} 
\label{fig:conditional_variability}
\end{figure}

\begin{figure*}[ht!]
\centering
\includegraphics[width=\textwidth]{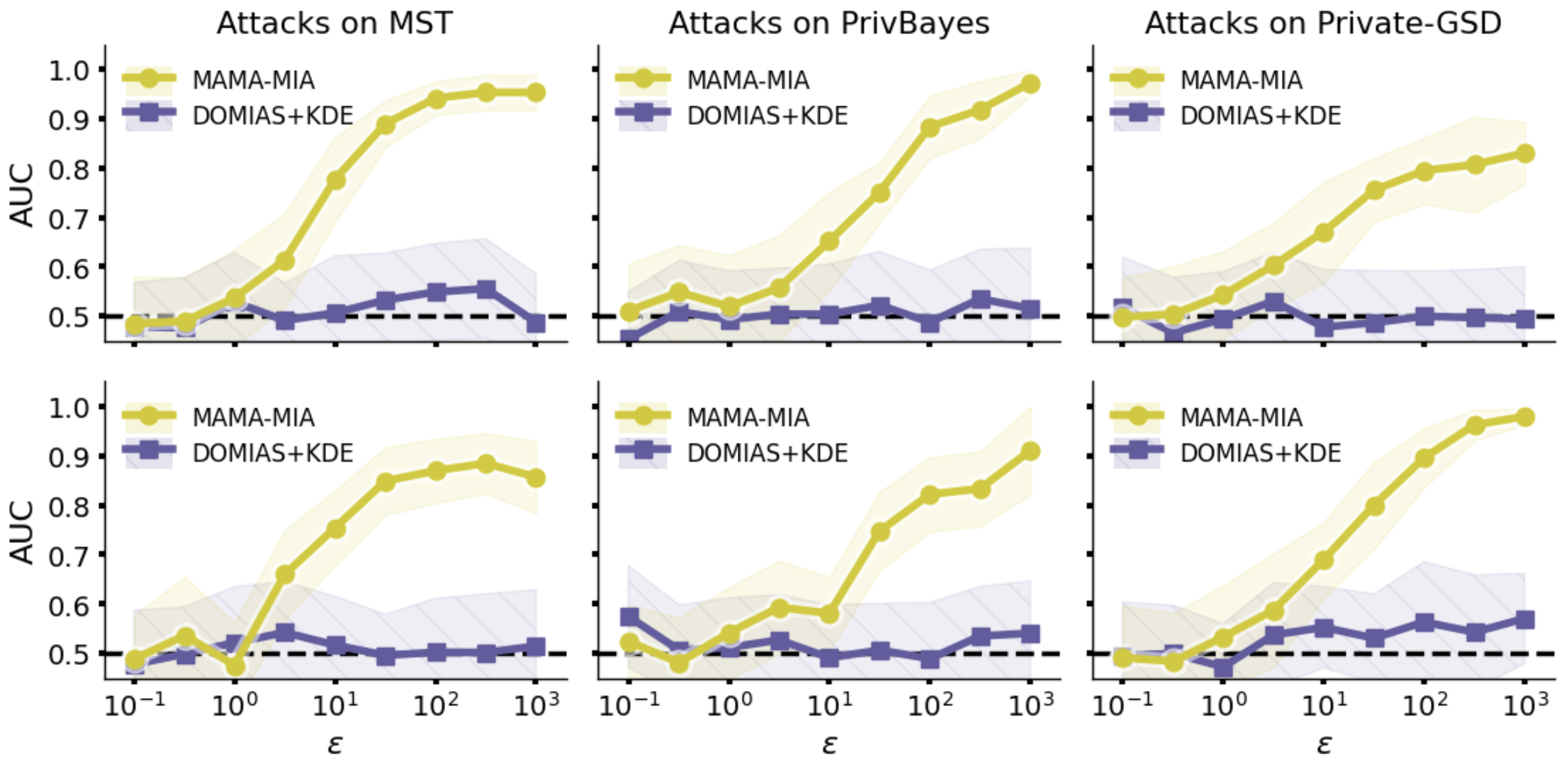}
\caption{MAMA-MIA and DOMIAS-using-KDE accuracies when performing set MI on the SNAKE data (top) and on the housing data (bottom). $|D_{target}| = 1,000$ records, $|D_{target}| = 32$ sets, and $|D_{member}| = 16$ sets.} 
\label{fig:setMI_snake}
\end{figure*}

\vspace{7pt}\noindent{\textbf{The importance of the hyperparameters in FP prediction}}
Over the course of our lengthy investigation, it became clear that predicting focal-points in any of the generators was folly if the wrong parameters were used in shadow modelling. This is especially true for choosing the correct $\varepsilon$, since the FP selection behavior changes in the SDG algorithms for different values of it.

As expected, we observed that the variability increased as the privacy-loss budget $\varepsilon$ decreased, because a smaller $\varepsilon$ amounts to a higher amount of randomness in the selection of focal-points by the SDG. For MST, we visualize these findings as a bar graph in Figure \ref{fig:marginal_variability}, where each bar represents a percentage of shadow runs FPs were reselected, across different values of $\varepsilon$. Notice that, when $\varepsilon = 0.1$ most of the marginals  selected by MST were only chosen less than $50\%$ of the time, boding poorly for our confidence in which FPs to measure. On the other hand, with higher $\varepsilon$, a vast majority of marginals were chosen more than $75\%$ of the time during shadow modelling, which makes our density estimation of $D_{synth}$ closer to that of the hidden $D_{train}$.

Moreover, PrivBayes \textit{preferred} different FPs for different $\varepsilon$. Figure \ref{fig:conditional_variability} shows a histogram of how frequent conditionals' parents sizes were selected, and the trends for different $\varepsilon$. When $\varepsilon = 0.1$, PrivBayes only ever allows for conditionals to have one or zero parents. But when $\varepsilon = 1000$, its graphical estimation is constructed by favoring conditionals with two, three, and up to four parents. This was expected because of the way PrivBayes changes its maximum parent size to make the most effective use of its privacy budget.

\vspace{7pt}\noindent{\textbf{Set MI experiments}}
We show results for set MI experiments in Figure \ref{fig:setMI_snake}. These are conducted using the same settings as in experiment A, i.e. with $|D_{train}|$ fixed at 1,000 records, trialling both the SNAKE and the housing data. We only present results for DOMIAS, when using KDE, and we omit set MI results for the attack from the TAPAS library because it does not offer set MI in its API. Sets are groups of at least five target records, and instead of making $|D_{target}|$, and $|D_{member}|$ equal to 32 and 16 records, respectively, we use 32 and 16 \textit{sets}. The results are unsurprising, where MAMA-MIA's accuracy improves over the single MI setting.

\vspace{7pt}\noindent{\textbf{MAMA-MIA's performance on the wrong synthetic data generator}}
Table \ref{tab:wrong_attacks} shows the interesting case when MAMA-MIA is tailored to one SDG algorithm, say MST, and is used perform membership inference on data produced by a different SDG algorithm. We show all combinations of tailoring MAMA-MIA to one of MST, PrivBayes, and Private-GSD, and using it to attack data from each of those SDGs. The columns depict which algorithm generated the data, and the rows depict which algorithm MAMA-MIA was tailored to. Naturally, the diagonal (in boldface) represents the case where MAMA-MIA attacks the correct data. This scenerio is carried out on the SNAKE data, when $\varepsilon = 10$ and $|D_{train}| = 1,000$.

These results are interesting because they show that MAMA-MIA performs better than random guessing when the data is not from the anticipated generator. This is despite using focal-points from an entirely different SDG algorithm in our construction of $\zeta$. However, as expected, MAMA-MIA attacking the data for which it is tailored produces far better results, demonstrating the advantage MAMA-MIA has with knowledge of which algorithm was used.

\begingroup

\setlength{\tabcolsep}{4pt} 
\begin{table}[ht!]
\caption{AUC scores $\pm$ one standard deviation, using MAMA-MIA tailored to one SDG algorithm to perform inference on synthetic data produced by another algorithm. $\varepsilon = 10$, using SNAKE data, and $|D_{train}| = 1,000$. The diagonal values (in boldface) are when MAMA-MIA attacks data generated by the intended SDG algorithm. }
\centering
\begin{tabular}[b]{c | c  c  c}
\multicolumn{1}{c}{Attack method} &\multicolumn{3}{c}{Data generated by} \\
      & MST & PrivBayes & Private-GSD \\ \hline
    MAMA-MIA & \multirow{2}{*}{$\mathbf{0.614 \pm .11}$} & \multirow{2}{*}{$0.504 \pm .13$} & \multirow{2}{*}{$0.523 \pm .10$} \\
    MST &  &  &  \\ \hline
    MAMA-MIA & \multirow{2}{*}{$0.539 \pm .11$} & \multirow{2}{*}{$\mathbf{0.578 \pm .09}$} & \multirow{2}{*}{$0.545 \pm .08$} \\
    PrivBayes &  &  &  \\ \hline
    MAMA-MIA & \multirow{2}{*}{$0.546 \pm .11$} & \multirow{2}{*}{$0.538 \pm .10$} & \multirow{2}{*}{$\mathbf{0.612 \pm .10}$} \\
    Private-GSD &  &  &  \\ \hline
   
\end{tabular}
\label{tab:wrong_attacks}
\end{table}

\endgroup

\subsection{Description of related MIAs}\label{sec:related_MIAs}

\textbf{Groundhog attack}\cite{stadler2022synthetic}: Stadler et al. assess the privacy gain of different SDGs by proposing targeted attacks. They consider a scenario in which only one copy of the generated synthetic data is published. They describe their threat model as a \textit{generic black-box MIA}. Specifically, they consider an adversary who has prior knowledge of the SDG (referred to as black-box knowledge in our paper) and access to a reference dataset (referred to as auxiliary data in our paper) that shares the same distribution as the training data. The underlying intuition of their attack is that if the ``real training data'' contains certain attributes, the synthetic data is likely to preserve these attributes, such as correlations and statistical metrics, even if the SDG model does not explicitly capture them.

To perform this attack, the adversary employs a shadow modeling strategy following Shokri et al. \cite{shokri2017membership}. The core idea is to sample a large number of training datasets from the auxiliary dataset, which are then subsequently and independently used to generate synthetic datasets. The goal of the adversary is then to infer if a specific target $t$ was included in the training set used to generate the synthetic dataset being attacked. To achieve this, Stadler et al. propose to train a Random Forest classifier with a rich set of input features -- (a) naive features such as median, mean and variance of each column in tabular dataset, (b) histogram based features that represent distributions of each column in tabular dataset, and (c) correlation features that capture the correlation coefficients. For further details, please refer to \cite{stadler2022synthetic} and its implementation in TAPAS\footnote{\url{https://tapas-privacy.readthedocs.io/en/latest/library-of-attacks.html}}


\vspace{0.8em}
\noindent
\textbf{TAPAS} \cite{houssiau2022tapas}:  TAPAS is a comprehensive framework comprising various types of attacks to evaluate the privacy of generated synthetic data across different scenarios, including different threat models and privacy evaluation metrics. In their paper, Houssiau et al. make novel tweaks to the GroundHog attack above to improve the accuracy of these attacks. Specifically, they adjust the set of features provided to the Random Forest Classifier. They randomly select $x$ subsets of $m$-ordered column feature sets in the tabular data. For each subset, they then compute the number of records in the synthetic data that match the target record $t$ on the selected column feature sets. These $x$ queries form the dimensions of the resulting feature vector that is fed to the Random Forest Classifier \footnote{We implemented this improved attack using the \texttt{RandomTargetedQueryFeature} in TAPAS library and extended it to consider a set of target records rather than a single target}. 


\vspace{0.8em}
\noindent
\textbf{DOMIAS} \cite{van2023membership}: DOMIAS proposed a MIA on synthetic data that leverages density estimation to detect if the generator was overfit on the training dataset. The authors assume a modified black-box setting where the adversary has access to one copy of synthetic dataset, a target record $t$ and additional access to a reference dataset (auxiliary dataset in our paper).  DOMIAS constructs the density estimators  on the reference ${p}_{ref}()$ and the synthetic ${p}_{syn}()$ datasets independently and then follows the equation below for scoring the MIA for target $t$
\begin{equation*}
    \mathcal{A_{DOMIAS}} = \frac{{p_{syn}(t)}}{{p_{ref}(t)}}
\end{equation*}
The intuition is similar to what we explain in section \ref{sec:method} and Figure \ref{fig:domias_example}.

\subsection{Limitations}

The generators studied in this work add DP noise on the measurements taken on the training data. So their probability estimations of $D_{train}$ will be quite inaccurate when $\varepsilon$ is small, thereby creating a $D_{synth}$ that inherently reveals little about $D_{train}$, making membership inference attacks very challenging. Also, when an SDG algorithm produces data that has inherently lower quality (such as in RAP's case), there is a limit to the degree MAMA-MIA can glean anything about the hidden data.

Clearly, MAMA-MIA achieves success by using a threat model, wherein an adversary is assumed access to the auxiliary data and black-box knowledge of the generator. Without these two assumptions, our attack is not applicable. Additionally, our experiments only include the setting where $|D_{member}| = 1/2 \cdot |D_{target}|$. Future work would include experiments where this proportion is varied, which may entail adjusting the activation function from section \ref{sec:activation}.

Also note that, even when $\varepsilon$ is arbitrarily large, and the added noise is negligible (or if granted white-box access), MAMA-MIA will not necessarily score perfect accuracy; this is because the generator builds only an \textit{estimation} of the true joint probability distribution. Only constructing a perfect probability distribution will yield perfect information of $D_{train}$, which is computationally infeasable when $D_{train}$ is even moderately sized.

As far as computational limitations, our workload is divided between calculating $\zeta$, which runs in linear time, respective to the size of the auxiliary dataset, and shadow modelling, which consists of only 50 trials. However, each trial requires executing the SDG algorithm, and so the SDG algorithm itself is a limiting factor. (In the case of Private-GSD, our ability to terminate the SDG algorithm early after it determines its focal-points shortens this process substantially.)

We do not study the ability of our attack for underrepresented groups in the data or outlying records since we choose our targets through random sampling.

As always, our attack is academically-motivated, with the hope that our strong results will drive the effort to develop stronger SDG protections.

\newpage
\onecolumn


\end{document}